\documentclass{article}
\usepackage{graphicx}  
\usepackage{amsmath}   
\usepackage[compress,numbers,sort]{natbib}
\usepackage{amssymb}   
\usepackage{bm} 
\usepackage{dcolumn}
\usepackage{color}
\usepackage{mathrsfs}
\usepackage{amsfonts}
\usepackage{varioref}
\usepackage{textcomp}
\usepackage[normalem]{ulem}

\usepackage{multirow}
\usepackage{subcaption}

\RequirePackage[colorlinks,citecolor=blue,urlcolor=magenta,linkcolor=blue]{hyperref}
\allowdisplaybreaks
\usepackage{orcidlink}
\labelformat{section}{Section #1} 
\labelformat{subsection}{Section #1} 
\labelformat{subsubsection}{Section #1}
\labelformat{subsubsubsection}{Section #1}
\labelformat{equation}{Eq.~(#1)} 
\labelformat{figure}{Fig.~#1} 
\labelformat{subfigure}{Fig.~\thefigure#1} 
\labelformat{table}{Table~#1} 
\labelformat{appendix}{Appendix #1}
\title{\bf Response of a Kerr black hole to a generic tidal perturbation}
\author{Rajendra Prasad Bhatt\orcidlink{0009-0004-9088-2998},$^{1,}$\footnote{\href{mailto:rajendra@iucaa.in}{rajendra@iucaa.in}} ~Sumanta Chakraborty\orcidlink{0000-0003-3343-3227},$^{2,}$\footnote{\href{mailto:tpsc@iacs.res.in}{tpsc@iacs.res.in}} ~and Sukanta Bose\orcidlink{0000-0002-4151-1347}$^{3,}$\footnote{\href{mailto:sukanta@wsu.edu}{sukanta@wsu.edu}}\\
$^{1}${\small{Inter-University Centre for Astronomy and Astrophysics, Pune 411007, India}}\\
$^{2}${\small{School of Physical Sciences}}\\
{\small{Indian Association for the Cultivation of Science, Kolkata-700032, India}}\\
$^{3}${\small{Department of Physics and Astronomy,}}\\
{\small{ Washington State University, 1245 Webster, Pullman, Washington 99164-2814, USA}}
}
\begin{document}
  
\maketitle
\begin{abstract} 
We derive the response, the real part of which provides the tidal Love numbers, for non-extremal as well as extremal Kerr black holes under generic tidal perturbations. Our results suggest that the static as well as dynamical (linear-in-frequency) Love numbers vanish for both Schwarzschild and slowly rotating (linear-in-angular momentum) Kerr black holes, under generic perturbations. The vanishing of static and dynamical Love numbers also holds for axisymmetric tidal perturbations of non-extremal and extremal Kerr black holes. In fact, even under generic tidal perturbations, the static Love numbers of Kerr black holes vanish identically. The only case with non-zero Love numbers corresponds to the non-axisymmetric dynamical tidal perturbations of Kerr black holes (to arbitrary order of angular momentum). We also demonstrate that the non-zero dynamical Love numbers, for both non-extremal and extremal Kerr black holes, get modified under the change of the sign of the spin-weight, for electromagnetic and gravitational tidal perturbations.
\end{abstract}

\section{Introduction}
Black holes (BHs), despite being the simplest and the most compact 
objects predicted by Einstein's general theory of relativity, pose various intriguing questions. The debate over these questions is further fueled by the recent observations of gravitational waves from  binary BH mergers~\cite{LIGOScientific:2016aoc, KAGRA:2021vkt} and the images of the shadows cast by  supermassive BHs~\cite{EventHorizonTelescope:2019dse, EventHorizonTelescope:2022wkp}. 
These observations have enabled us to study these fascinating objects in 
unprecedented detail 
and have also launched further theoretical investigations into the nature of BHs, with the well-founded hope that some of them would be 
supported observationally 
by more sensitive observatories in the near future. 
One such observable, over which there has been much debate recently, is the response of a BH to an external tidal field. To shed more light on this debate, in this work, we provide a comprehensive analysis
of the response function of an arbitrarily rotating BH subjected to various types of perturbations, such as scalar, electromagnetic, and gravitational.
This study is an extension of our previous works~\cite{Bhatt:2023zsy, Bhatt:2024yyz}, where we analyzed the response of a BH to the gravitational tidal field of spin-weight $s = -2$ alone.

Response of a self-gravitating body to an external tidal field can be classified into two categories: one is the deformation caused by the tidal field (conservative), and the other is the loss of energy and angular momentum due to tidal effects (dissipative)~\cite{Chia:2020yla}. 
The conservative and dissipative parts of the tidal response can be quantitatively studied by dimensionless numbers known as tidal Love numbers (TLNs) and tidal dissipation numbers, respectively~\cite{Hinderer:2007mb, Flanagan:2007ix, Damour:2009vw, Binnington:2009bb, LeTiec:2020spy, LeTiec:2020bos, Chia:2020yla, Bhatt:2023zsy}.
The study of tidal response is usually done for two scenarios: static and dynamic. In the static case, the self-gravitating body and tidal fields are held fixed in the reference frame of an external observer, while in the dynamic case, the self-gravitating body and tidal fields move relative to each other. The tidal response function associated with these tidal fields are named as static and dynamic tidal response functions, respectively. Similar nomenclature holds for the tidal Love numbers and the tidal dissipation numbers. At this outset, we would like to emphasize that the word ``tidal'' in both the tidal response function as well as the tidal Love number generally refers to the computations involving gravitational perturbation. However, in order to have a uniformity, we will refer to the response function and Love numbers of scalar and electromagnetic perturbations also with the word ``tidal''. Since the nature of the perturbation will be clear from the context, we will keep using ``tidal'' for brevity.

Various studies have been performed to calculate the TLNs of black holes. Static TLNs of non-rotating and rotating black holes have been found to be zero in various studies~\cite{Binnington:2009bb, Damour:2009vw, Kol:2011vg, Chakrabarti:2013lua, Gurlebeck:2015xpa, Landry:2015zfa, Pani:2015hfa, LeTiec:2020spy, LeTiec:2020bos, Chia:2020yla, Charalambous:2021mea, Hui:2020xxx, Creci:2021rkz, Bhatt:2023zsy, Chia:2023tle, Bhatt:2024yyz, Kehagias:2024rtz, Gounis:2024hcm, Sharma:2024hlz}. However, the dynamical TLNs of non-extremal and extremal rotating black holes are found to be non-vanishing for non-axisymmetric gravitational tidal field with spin-weight $s=-2$~\cite{Bhatt:2024yyz, Chakrabarti:2013lua, Saketh:2023bul, Perry:2023wmm, Perry:2024vwz}. In addition, TLNs are also studied for neutron stars~\cite{Hinderer:2007mb, Damour:2009vw, Landry:2015zfa, Pani:2018inf, Landry:2015cva, Lowrey:2024anh, Creci:2023cfx, Yagi:2013bca, HegadeKR:2024agt, Chakravarti:2019aup}, ultra compact objects~\cite{Cardoso:2017cfl, Mendes:2016vdr, Raposo:2018rjn, Pani:2015tga, Uchikata:2016qku, Chakraborty:2023zed, Nair:2022xfm}, BHs in alternative as well as higher dimensional theories of gravity~\cite{DeLuca:2022tkm, Charalambous:2023jgq, Rodriguez:2023xjd, Chakravarti:2018vlt}, BHs in asymptotically non-flat spacetimes~\cite{Nair:2024mya, Emparan:2017qxd, Franzin:2024cah}, covariant loop quantum BHs~\cite{Motaharfar:2025ihv}, and Ba\~nados-Teitelboim-Zanelli (BTZ) black holes~\cite{DeLuca:2024ufn, Bhatt:2024mvr}. There also 
exist 
results involving TLNs for BHs 
immersed in an environment made up of dark matter or accretion \cite{Cannizzaro:2024fpz, DeLuca:2024uju}. Additionally, TLNs have been used to explore the quantum gravitational effects~\cite{Motaharfar:2025typ, Motaharfar:2025izo} and implications on extreme mass ratio binary system~\cite{Camilloni:2023rra, Grilli:2024fds}.

From post-Newtonian theory to black hole perturbation theory to effective field theory (EFT), various ways to calculate the TLNs of a black hole have been proposed. However, some of the past approaches  suffered from 
the presence of  ambiguities 
in their calculations~\cite{Gralla:2017djj}. In Refs.~\cite{LeTiec:2020spy, LeTiec:2020bos}, the authors proposed a procedure to deduce the tidal response function of a black hole from the calculation of the Weyl scalar, where they derived the Weyl scalar from the Teukolsky equation (i.e., the equation governing the perturbation of Kerr black holes). They also reported non-zero but imaginary static response function for Kerr black holes in non-axisymmetric tidal background. Since the calculation performed in Refs.~\cite{LeTiec:2020spy, LeTiec:2020bos} was done for static tidal fields, Ref.~\cite{Chia:2020yla} extended the calculation and argued that the non-zero response functions calculated in Refs.~\cite{LeTiec:2020spy, LeTiec:2020bos} are not associated with the tidal deformation but with tidal dissipation. In Ref.~\cite{Bhatt:2023zsy}, we showed that some terms were missing in the approximated Teukolsky equation used in Ref.~\cite{Chia:2020yla} and calculated the tidal response function for Schwarzschild and slowly rotating Kerr black hole. 
In Ref.~\cite{Bhatt:2024yyz}, we extended that study and calculated the tidal response function of arbitrarily  rotating black holes, both
non-extremal and extremal. We showed that the dynamical tidal Love numbers of a rotating black hole can be non-zero, in general. This work was done for gravitational perturbation with spin-weight $s=-2$. In this work, we extend the analysis of Refs.~\cite{Bhatt:2023zsy, Bhatt:2024yyz}, to calculate the tidal response function of arbitrarily  rotating black holes (non-extremal and extremal) for scalar ($s=0$), electromagnetic ($s=\pm1$), and gravitational ($s=\pm2$) perturbations. Our results are valid for the perturbations defined by integer spin-weight $s$~\cite{Teukolsky:1972my, Teukolsky:1973ha, Press:1973zz, Teukolsky:1974yv}.

This article is arranged as follows. In \ref{section:2}, we shall briefly review the definition of the tidal response function and the procedure used to calculate it. In \ref{sec:nextkerr}, we will study the tidal response of a non-extremal rotating BH. To do so, we will present the Teukolsky equation in small frequency and near-zone regime. In \ref{tide_scalar}, we will study the response of a non-extremal rotating BH to the scalar tidal field. In \ref{sec:Tidal_response_ext_rot_BH_cal}, we will calculate the response of an extremal rotating black hole to the external tidal field. Finally, in \ref{section:6} we will summarize the results found in this work and discuss their implications and future avenues of research. Some derivations are detailed in the appendices.

{\em Notation and conventions} --- We set $G=c=1$, unless otherwise stated. We work with the positive signature metric; i.e., the Minkowski metric in Cartesian coordinate system will be given by $\text{diag }(-1,+1,+1,+1)$. $\mathbb{Z}^{+}$ or $\mathbb{Z}_{\ge0}$ defines the set of non-negative integers $\{0, 1, 2, 3,\,.\,.\,.\}$.

\section{Response of an object immersed in an external tidal field}\label{section:2}

We begin with a concise introduction to the response function of a self-gravitating body in an external tidal field. This scenario can be realized in any binary system--from the Earth-Moon system to the extreme case of a binary BH inspiral \cite{poisson_will_2014, LeTiec:2020bos, Bhatt:2023zsy}.
Even 
in the absence of an external tidal field, in general an object would possess a definite multipolar structure to begin with. In the case of spherical symmetry, only the monopole moment is present, while in the more complicated case of axial symmetry (e.g., the Kerr BH), higher moments are also present, but they are all determined in terms of intrinsic properties of the body (e.g., mass, angular momentum, etc.). When these bodies are immersed in an external tidal field, which can have scalar, electromagnetic, and gravitational origin, the bodies are deformed, i.e., their multipole moments change. This change in their multipole moments is what is referred to as the tidal response of the self-gravitating object under the external tidal field.     

The notion of tidal response is most straightforward in the Newtonian context with a spherically symmetric non-rotating body of mass $M$ placed in an external tidal field. The Newtonian potential at a position $r$ from the center of mass of the body will be the linear combination of the potential due to the body as well as the potential created by the external tidal field, 
such that, the total potential at a given spatial point $r$ reads~\cite{LeTiec:2020bos, poisson_will_2014, Bhatt:2023zsy}
\begin{equation}
U=U_{\text{body}}+U_{\text{tidal}} = \frac{M}{r} + \sum_{l=2}^{\infty}\sum_{m=-l}^{l}\left[\frac{(2l-1)!!}{l!}\frac{I_{lm}Y_{lm}}{r^{l+1}}-\frac{(l-2)!}{l!}\mathcal{E}_{lm}Y_{lm}r^l\right]~,
\label{total_potential}
\end{equation}
where $(M/r)$ is the monopole moment of the compact object, describing the potential in absence of the external tidal field. Owing to spherical symmetry of the problem, any field can be decomposed into spherical harmonics, $\mathcal{E}_{lm}$ are the spherical components of the tidal field, and $I_{lm}$ are the spherical components of the induced multipole moment on the body~\cite{LeTiec:2020bos, poisson_will_2014, Bhatt:2023zsy}. 
Since the multipole moments $I_{lm}$ are induced on the body, due to the external tidal field $\mathcal{E}_{lm}$, it follows that, at least in the linear approximation, these two are proportional to each other~\cite{LeTiec:2020bos, Chia:2020yla, Bhatt:2023zsy},
\begin{equation}
I_{l m} = -\frac{(l-2)!}{(2l-1)!!} R^{2l+1} \sum_{l' m'}[2 k _{l m; l' m'}\mathcal{E}_{l' m'} -\tau_0 \nu_{l m; l' m'} \dot{\mathcal{E}}_{l' m'} + \ldots]~.
\end{equation}
This expression connecting multipole moments and the tidal field holds for any spinning object, which generically involves coupling between different multipole indices $l$ and azimuthal indices $m$. Since our focus in this work is in Kerr black hole, it follows that such mode mixing does not arise (for more details, see Ref.~\cite{LeTiec:2020bos}). Thus, the above equation connecting multipoles with tides simply becomes
\begin{equation}
I_{lm} = -\frac{(l-2)!}{(2l-1)!!} R^{2l+1} \left[2k_{lm} \mathcal{E}_{lm} -\tau_0 \nu_{lm} \dot{\mathcal{E}}_{lm} + \cdots\right]~.
\label{I_def}
\end{equation}
Here, $k_{lm}$ are the TLNs, and $\nu_{lm}$ are the tidal dissipation numbers, both dimensionless, with $\tau_0$ being the viscosity induced time delay~\cite{Chia:2020yla, Bhatt:2023zsy}. From dimensional grounds, it follows that the relation between $I_{lm}$ and $\mathcal{E}_{lm}$ must have a term with dimension $[\textrm{Length}]^{2l+1}$, which is achieved by introducing the $R^{2l+1}$ factor, where $R$ is related to the size of the body. Note that the overall normalization factor $\{-(l-2)!/(2l-1)!!\}$ might be different in other studies, depending on the conventions employed. 

Using the relation between $I_{lm}$ and $\mathcal{E}_{lm}$, in the Fourier space, we can write down the total Newtonian potential from \ref{total_potential} as
\begin{equation}
U=\frac{M}{r}-\sum_{l=2}^{\infty}\sum_{m=-l}^{l}\frac{(l-2)!}{l!}\mathcal{E}_{lm}\,r^l\left[1+F_{lm}(\omega)\left(\frac{R}{r}\right)^{2l+1}\right]Y_{lm}~,
\label{total_potential_Fourier_space}
\end{equation}
where $F_{lm}\equiv 2k_{lm}+i\omega\tau_0 \nu_{lm}+\mathcal{O}(\omega^{2})$ is defined as the tidal response function with $\omega$ being the mode frequency. Even though the above appears to provide a complete picture, it is limited to the Newtonian context. Since we wish to understand the tidal response function in the context of BH spacetimes, it requires the introduction of the tidal response function in the relativistic scenario.   

In the relativistic context, instead of the potential we must work with a gauge/diffeomorphism invariant quantity, e.g., a scalar. Moreover, this scalar function, in the non-relativistic limit, must act as a proxy for the Newtonian potential, helping us in connecting the relativistic tidal response function to its non-relativistic counterpart. As discussed in Refs.~\cite{LeTiec:2020bos, Bhatt:2023zsy}, Newman-Penrose scalars, e.g., $\Psi_{4}$~\cite{Teukolsky:1972my, Teukolsky:1973ha}, are suitable for capturing tidal effects in the relativistic context. In the Newtonian limit
\begin{equation}\label{calc_newtonian_psi4_1}
\lim_{c\rightarrow\infty}c^2\Psi_4=-\frac{1}{2}\Bar{m}^i\Bar{m}^j\nabla_i\nabla_jU\,.
\end{equation}
Here, $\nabla_i$ is a covariant derivative associated with the three dimensional Euclidean metric in the Cartesian coordinates, $\bar{m}^{i}$ is a complex Newman-Penrose vector on the two sphere, and $U$ is the Newtonian potential defined in \ref{total_potential}. In particular, the above expression for the Newman-Penrose scalar $\Psi_4$ follows from the following structure for the Newtonian metric, $g_{00} = -1 +(2U/c^2)+\mathcal{O}(c^{-4})$, $g_{0i} = \mathcal{O}(c^{-3})$, and $g_{ij} = \delta_{ij} [1 +(2U/c^2)] + \mathcal{O}(c^{-4})$. Further, the expression of $\bar{m}^{i}$ in the spherical polar coordinates is $(1/\sqrt{2}r)(\partial_\theta - i\,\csc\theta\partial_\phi)$. All of this ensures that indeed the Newman-Penrose scalars are the appropriate ones to determine the tidal response function of a compact object in a relativistic context. Thus, with \ref{total_potential_Fourier_space} and \ref{calc_newtonian_psi4_1}, we arrive at the following Newtonian limit of $\Psi_{4}$ (for more details about this calculation, see Refs.~\cite{LeTiec:2020bos, Bhatt:2023zsy}):
\begin{equation}\label{calc_newtonian_psi4}
\lim_{c\rightarrow\infty}c^2\Psi_4=\lim_{c\rightarrow\infty}c^2\sum_{lm}\Psi_4^{lm}=\sum_{lm}\frac{1}{4}\sqrt{\frac{(l+2)(l+1)}{l(l-1)}}\mathcal{E}_{lm}\,r^{l-2}\left[1+F_{lm}\left(\frac{R}{r}\right)^{2l+1}\right]\,_{-2}Y_{lm}~.
\end{equation}
 Unlike the Newtonian case, in the relativistic context, tidal fields are not only of gravitational origin but can also be due to an external scalar or electromagnetic field. There are Newman-Penrose scalars associated with these perturbations as well and are denoted by $\zeta^{(s)}$. This can be connected to the standard convention, as in Refs.~\cite{Teukolsky:1972my, Teukolsky:1973ha}, in the following manner: (a) for scalar perturbation ($s=0$), we have $\zeta^{(0)}=\Phi$; (b) for electromagnetic perturbation ($s=\pm 1$), we have $\zeta^{(1)}=\phi_{0}$, and $\zeta^{(-1)}=\phi_{2}$; and finally (c) for gravitational perturbation ($s=\pm 2$), we have $\zeta^{(2)}=\Psi_{0}$, and $\zeta^{(-2)}=\Psi_{4}$.  

Therefore, to calculate the response function under spin-$s$ external tidal perturbation, our main goal would be to calculate the corresponding Newman-Penrose scalar, by solving the Teukolsky equation (described in the next section). Once we have the solution for the Newman-Penrose scalar, we can find its large $r$ behavior (or behavior in the intermediate region), which will be similar to \ref{calc_newtonian_psi4}; i.e., it will have a part growing with the radial coordinate $r$, and another part decaying with $r$. The coefficient of the decaying part provides the tidal response function. To simplify and determine the response function in terms of BH hairs, we need to work with analytically continued angular number $l$, i.e., $l\in\mathbb{C}$~\cite{LeTiec:2020bos}. At the end of the calculation, to get the physical tidal response function, we must set $l\in\mathbb{Z}_{\ge0}$ and $s\in\mathbb{Z}$, such that $l\ge|s|$. The above analytic continuation scheme for TLNs is absolutely essential, since this decouples the post-Newtonian corrections from the actual tidal effects. This is because the $n$-th order post Newtonian term contributes as $(v^{2})^{n}\sim r^{-n}$, which can be unambiguously separated from the tidal response by taking $l$ to the complex domain~\cite{LeTiec:2020bos}.

Given this response function for the compact object, the tidal Love numbers and the tidal dissipation numbers,  under different types of perturbations, can be determined through the following definition:
\begin{equation}
\,_{s}k_{lm}\equiv\frac{1}{2}\text{Re}\,_{s}F_{lm}, \qquad 
\omega\tau_0\,_{s}\nu_{lm}\equiv\text{Im}\,_{s}F_{lm}\,.
\end{equation}
This is because the tidal Love numbers capture conservative part of the dynamics, while the dissipation numbers are associated with the dissipative parts. In what follows, we will solve the Teukolsky equation for generic spin perturbation and hence determine the tidal response function, whose real part provides the dynamical tidal Love numbers under generic spin perturbation of Kerr BH. 

\section{Response of a non-extremal Kerr black hole to generic tidal perturbations}\label{sec:nextkerr}

Having outlined the main procedure for calculating the response function of a compact object to the scalar, electromagnetic, and gravitational tidal perturbations from the corresponding Newman-Penrose scalars $\zeta^{(s)}$, we will apply this formalism to compute the dynamical tidal response of a non-extremal Kerr BH under generic perturbations. For this purpose, we present below the Teukolsky equations, corresponding to generic perturbations of the Kerr BH and then solve it in the near-zone and small frequency approximations. From the asymptotic expansion of this near-zone solution, we identify the tidal response function and hence determine the TLNs in a dynamical context.
\subsection{Teukolsky equation for arbitrary spin in the near-zone and small-frequency limit}

In this section, we will be studying the tidal response of a Kerr BH under generic spin perturbations, and these are described by the corresponding Teukolsky equations. Note that the Teukolsky equations are not for the Newman-Penrose scalars $\zeta^{(s)}$ but for the radial and angular parts of the combination $\rho^{-s+|s|}\zeta^{(s)}$, where $\rho=-(r-ia \cos\theta)$ is one of the spin coefficients. The above combination for different spin cases has been summarized in \ref{table_1}. 
\begin{table}[ht]
    \centering
    \setlength{\tabcolsep}{20pt}
    \renewcommand{\arraystretch}{1.5}
    \begin{tabular}{ccc}
    \hline
        Type of perturbation & $s$ & $\rho^{-s+|s|} \zeta^{(s)}$ \\ \hline
        Scalar &  $~~0$ & $\Phi$ \\
        
        Electromagnetic &  $~~1$ & $\phi_0 \equiv F_{\mu\nu}m^\mu l^\nu$ \\ 
        Electromagnetic &  $-1$ & $\rho\phi_2 \equiv \rho F_{\mu\nu}n^\mu \Bar{m}^\nu$ \\
        Gravitational &  $~~2$ & $\Psi_0 = C_{\mu \nu \alpha \beta}l^\mu m^\nu l^\alpha m^\beta$ \\
        Gravitational &  $-2$ & $\rho^4\Psi_4 = \rho^4 C_{\mu \nu \alpha \beta}n^\mu \Bar{m}^\nu n^\alpha \Bar{m}^\beta$\\ \hline
    \end{tabular}
\caption{Here we present the explicit forms of the field quantity $\rho^{-s+|s|} \zeta^{(s)}$ for different spin-$s$ perturbations, with $\rho\equiv-(r-ia\cos\theta)$ as one of the spin coefficients. We present the explicit forms of this quantity for scalar ($s=0$), electromagnetic $(s=\pm 1)$, and gravitational ($s=\pm 2$) perturbations and relate them to the scalar field $\Phi$, Maxwell stress tensor $F_{\mu \nu}$, and the Weyl tensor $C_{\mu \nu \alpha \beta}$, respectively~\cite{Teukolsky:1972my, Chandrasekhar:1985kt, DAmbrosio:2022clk}. Here, $\{l^\mu, n^\mu, m^\mu, \Bar{m}^\mu\}$ is a set of null tetrad vectors.}
\label{table_1}
\end{table}

For our purpose, it will be convenient to express the Teukolsky equation for generic spin perturbation in the ingoing null coordinate system: 
$\{v,r,\theta,\widetilde{\phi}\}$. Here, $r$ is the Boyer-Lindquist radial coordinate, with $dv\equiv dt+\{(r^{2}+a^{2})/\Delta\}dr$, and $d\widetilde{\phi}\equiv d\phi+(a/\Delta)dr$, where $a$ is the rotation parameter of the Kerr BH and $\Delta\equiv r^{2}-2Mr+a^{2}$.  
In these coordinates, we can decompose the combination $\rho^{-s+|s|} \zeta^{(s)}$ into radial and angular parts as~\cite{Teukolsky:1974yv}
\begin{equation}\label{gen_eq_decomp}
\rho^{-s+|s|} \zeta^{(s)}=\int \mathrm{d}\omega\,e^{-i\omega v}\sum_{lm} e^{-im\widetilde{\phi}}\,_{s}S_{lm}(\theta)\,_{s}R_{lm}(r)~,
\end{equation}
where $\rho\equiv -(r-ia\cos\theta)$ has been employed above, $\,_{s}S_{lm}(\theta)$ is the spin-weighted spheroidal harmonics and $\,_{s}R_{lm}(r)$ is the radial function, which is of our prime interest. The radial part of the spin-$s$ perturbation of the Newman-Penrose scalar; namely, $\,_{s}R_{lm}(r)$, satisfies the equation (known as the radial Teukolsky equation)~\cite{Teukolsky:1974yv}
\begin{equation}
\Delta\frac{\mathrm{d}^2\,_{s}R_{lm}}{\mathrm{d}r^2}+2\left[(s+1)(r-M)-iK\right]\frac{\mathrm{d}\,_{s}R_{lm}}{\mathrm{d}r}+\left[-\frac{4is(r-M)K}{\Delta}+2(2s-1)i\omega r-\lambda\right]\,_{s}R_{lm}=\,_{s}T_{lm}~,
\label{TEq}
\end{equation}
where $\,_{s}T_{lm}$ is the source term, $\Delta$ has been defined above, $K\equiv(r^{2}+a^{2})\omega-am$, and $\lambda\equiv E_{lm} -2am\omega +a^2\omega^2 -s(s+1)$, with the separation constant $E_{lm}$ having the following frequency expansion~\cite{Press:1973zz, Fackerell:1977,Seidel:1988ue, Berti:2005gp}
\begin{align}
E_{lm}&=l(l+1)+a^2\omega^2\left[\frac{2m^2-2l(l+1)+1}{(2l-1)(2l+3)}\right]+\mathcal{O}[(a\omega)^4],\qquad s=0
\\
&=l(l+1)-2a\omega\frac{s^2m}{l(l+1)}+\mathcal{O}[(a\omega)^2]~,\qquad \qquad \qquad \quad \quad \quad s\neq 0~.
\end{align}
Here, $M$ is the mass of the BH, and $J=aM$ is the angular momentum. Note that, in Ref.~\cite{Teukolsky:1974yv}, the term $\{-2(2s+1)i\omega r\}$ was written in place of $\{2(2s-1)i\omega r\}$ in \ref{TEq}, which was incorrect~\cite{Chia:2020yla}. In addition, the coefficient of $(a\omega)^2$ was also different in the expression of $E_{lm}$ for scalar perturbation ($s=0$) in Ref.~\cite{Press:1973zz}, which we have corrected here (see Refs.~\cite{Fackerell:1977, Seidel:1988ue, Berti:2005gp}). In addition, $l$ and $m$ are multipole and azimuthal indices (or angular and magnetic quantum numbers), respectively, where $l\in \mathbb{Z}^+$, i.e., $l$ is always a positive integer including zero, with $l\ge |s|$, and $m\in[-l,\,l]$, with $s,\,m\in \mathbb{Z}$. 

The above radial equation can be expressed in a more suggestive form by writing down $\Delta=(r-r_{+})(r-r_{-})$, where $r_{\pm} \equiv M\pm\sqrt{M^{2}-a^{2}}$ denote the radial positions of the event and the Cauchy horizons, respectively. Thus, we can rewrite the radial Teukolsky equation, presented in \ref{TEq}, as
\begin{multline}\label{Teqr}
\frac{\mathrm{d}^2\,_{s}R_{lm}}{\mathrm{d}r^2}+\left(\frac{2iP_++(s+1)}{r-r_+}-\frac{2iP_--(s+1)}{r-r_-}-2i\omega\right)\frac{\mathrm{d}\,_{s}R_{lm}}{\mathrm{d}r} 
\\ 
+\left[\frac{2isP_+}{(r-r_+)^2}-\frac{2isP_-}{(r-r_-)^2}+\frac{A_-}{(r-r_-)(r_+-r_-)}-\frac{A_+}{(r-r_+)(r_+-r_-)}\right]\,_{s}R_{lm}=\frac{\,_{s}T_{lm}}{\Delta}~,
\end{multline}
where we have introduced the quantities $P_{\pm}$ and $A_{\pm}$, having the following expressions:
\begin{equation}\label{def_Ppm}
P_\pm = \frac{am-2r_\pm M\omega}{r_+-r_-}~,\qquad A_\pm = 2i\omega r_\pm +\lambda~.
\end{equation}
If we now apply the following transformation: $z\equiv(r-r_+)/(r_+-r_-)$, which effectively shifts the origin of the coordinate system to the radial position of the event horizon, then the radial Teukolsky equation, as in \ref{Teqr}, becomes
\begin{multline}\label{TZ}
\frac{\mathrm{d}^2\,_{s}R_{lm}}{\mathrm{d}z^2} + \left[\frac{2iP_++(s+1)}{z}-\frac{2iP_--(s+1)}{1+z}-2i\omega(r_+-r_-)\right]\frac{\mathrm{d}\,_{s}R_{lm}}{\mathrm{d}z} 
\\ 
+\left[\frac{2isP_+}{z^2}-\frac{2isP_-}{(1+z)^2}+\frac{A_-}{1+z}-\frac{A_+}{z}\right]\,_{s}R_{lm}=\frac{\,_{s}T_{lm}}{\Delta}(r_+-r_-)^2~.
\end{multline}
To simplify the above equation, we will apply small frequency $(M\omega\ll1)$ and near-zone $(M\omega z\ll1)$ approximation.\footnote{Regarding the near-zone expansion, we would like to point out that ignoring $M\omega z$, but keeping $M\omega$ is a standard approximation, employed in connection with determining the dynamical response function~\cite{Chia:2020yla, Charalambous:2021mea, Hui:2022vbh}. Moreover, even the computation of quasi-normal modes using the matched asymptotic expansion scheme~\cite{Cardoso:2008kj} employ exactly the same approximation, which reproduces the mode frequencies correctly. The argument behind this approximation, at least for the computation of the TLNs is as follows: the determination of the TLNs require probing the intermediate region, which is away from the deformed object as well as the tidal field, but \emph{not} the asymptotic region. The intermediate regime is determined such that $M\omega z$ remains a very small quantity. Therefore, the assumption of ignoring $M\omega z$ is consistent, if we restrict our analysis to the intermediate zone alone, and not the asymptotic/far zone. Here, also we only refer to the intermediate zone and not to the asymptotic/far zone.} Note that in the context of effective field theory (EFT) approach as well, such a near-zone expansion is routinely used; see, for example, Ref.~\cite{Creci:2021rkz}. In the perturbation theory approach, one uses the asymptotic form of the near-zone approximation to determine the Love numbers, while in the EFT approach, one matches the near-zone solution around the compact object with asymptotic solution, which is then matched with the flat spacetime effective field theory involving point particle action along with various multipole moments. Thus, the near-zone approximation is routine in both the perturbation theory as well as the EFT approach. Moreover, such a scheme for computing the near-zone solution and then expanding in the intermediate regime to get the Love numbers can be ambiguous due to inability to properly segregate the tidal and the response part. This is achieved by invoking the analytic continuation of the angular number $l$; i.e., from this point onward, we will treat $l$ as a complex quantity and shall set it to be an integer at the very end. In particular, ignoring all second- and higher-order terms of $M\omega$, as well as setting the source term to zero, the above radial equation boils down to~\cite{Bhatt:2023zsy}
\begin{multline}\label{GMATE}
\frac{\mathrm{d}^2\,_{s}R_{lm}}{\mathrm{d}z^2} + \left[\frac{2iP_++(s+1)}{z} - \frac{2iP_1-(s+1)}{1+z}\right]\frac{\mathrm{d}\,_{s}R_{lm}}{\mathrm{d}z}  +\left[\frac{2isP_+}{z^2}-\frac{2isP_2}{(z+1)^2}\right.
\\
\left.-\frac{l(l+1)-s(s+1)}{z(1+z)}+\frac{2am\omega}{z(1+z)}\left\{1+\frac{s^2}{l(l+1)}\right\} - \frac{2i\omega r_+}{z(1+z)}\right]\,_{s}R_{lm}=0~.
\end{multline}
where we have defined the two constants $P_{1}$ and $P_{2}$ as 
\begin{equation}
P_1\equiv P_- + \omega (r_+-r_-)~,\qquad P_2\equiv P_-+\frac{1}{s}\omega(r_+-r_-)~.
\end{equation}
It is to be noted that \ref{GMATE} is valid for scalar, electromagnetic, and gravitational perturbations, and this will be our master equation for this work. However, the above equation does not work for extremal Kerr BH (for which $r_{+}=r_{-}$), due to the ill-behaved nature of the coordinate $z$, in the extremal limit. Thus, we will present the corresponding computation for an extremal Kerr BH, separately, in the next section. Therefore, having setup the relevant equation and the approximation scheme involved, we will find its solution with purely ingoing boundary condition at the horizon and hence will determine the dynamical TLNs associated with generic spin perturbation of the Kerr BH.
\subsection{Tidal response from the expansion of the Newman-Penrose scalar}\label{sec:Tidal_response_arb_rot_BH_cal}

The master equation derived in the last section is a second-order differential equation having regular singular point at $z = 0,\, -1,$ and $\infty$. Thus, the master equation can be solved in terms of the Gauss hypergeometric function~\cite{Arfken_Weber_2005, abramowitz_stegun_1972},
\begin{multline}\label{radialfngen_1}
\,_{s}R_{lm}(z)=(z+1)^{-N_3-s} \left[c_1 z^{-s} \, _2F_1\left(l-s-N_2+1,-l-s-N_1;-s+2 i P_++1;-z\right)\right.\\\left.+c_2 z^{-2 i P_+} \, _2F_1\left(l-2 i P_+-N_2+1,-l-2 i P_+-N_1;s-2 i P_++1;-z\right)\right],
\end{multline}
where $c_1$ and $c_2$ are the constants of integration. The arguments of hypergeometric functions and the power of $(1+z)$ are written up to the linear orders of $M\omega$, and we have introduced three linear-in-frequency quantities, $N_1$, $N_2$, and $N_3$, having the expressions
\begin{align}\label{def_N1}
N_{1}=2\omega\Big[-\frac{a m s^2}{l(l+1)(2l+1)}&+\frac{i \left(i a m+4 l M-2 l r_++r_+\right)}{2 l+1}
\nonumber
\\
&+\frac{2 \left(M-r_+\right) \left(-ia m+M-r_+\right)}{\left(a m+i M s-i r_+ s\right)}+\frac{2 i s \left(M-r_+\right)}{2 l+1}\Big]~,
\end{align}
\begin{align}\label{def_N2}
N_{2}=2\omega\Big[\frac{a m s^2}{l(l+1)(2l+1)}&+\frac{a m+4 i (l+1) M-i (2 l+3) r_+}{2 l+1}
\nonumber
\\
&+\frac{2 \left(M-r_+\right) \left(-i a m+M-r_+\right)}{\left(a m+i M s-i r_+ s\right)}-\frac{2 i s \left(M-r_+\right)}{2 l+1}\Big]~,
\end{align}
and,
\begin{equation}\label{def_N3}
N_{3}=\frac{4i\omega (s-1) \left(M-r_+\right){}^2}{\left(-i a m+M s-r_+ s\right)}~.
\end{equation}
Note that the above expressions of $N_1$, $N_2$, and $N_3$ are not well defined for Schwarzschild $(a=0)$ and axisymmetric tidal perturbation $(m=0)$ for scalar tidal field $(s=0)$. Hence, we will analyze the scalar case separately in \ref{tide_scalar}. Thus, the calculation presented in this section is only valid for $s\ne0$ cases. In addition, we would like to point out that
the above solution for the radial function $\,_{s}R_{lm}$ correctly reproduces the solution of Ref.~\cite{Bhatt:2024yyz} as well as the solution presented in Ref.~\cite{Bhatt:2023zsy} associated with Schwarzschild and slowly rotating Kerr BH, for $s=-2$, i.e., for gravitational perturbation.

Since our interest is in the determination of the response of the Kerr BH under external tidal perturbation, we will employ purely ingoing boundary condition at the horizon. 
To apply the above boundary condition, we provide below the behavior of the radial perturbation, as it approaches the horizon scale, i.e., in the limit $z\to 0$:
\begin{equation}\label{near_limit_1}
\,_{s}R_{lm}^{\rm near}(z)\sim c_1 z^{-s}+c_2 z^{-2 i P_+}\,.
\end{equation}
Among the two terms in the above solution, the $z^{-s}$ term arises from $\Delta^{-s}$, which depicts the purely ingoing term at the event horizon~\cite{Teukolsky:1973ha, Teukolsky:1974yv}, while the term $z^{-2 i P_+}$ can be written as $\exp(-2iP_{+}\ln{z})\approx \exp(2i\omega'r_{*})$, which is purely outgoing at the event horizon (here, $r_{*}$ is the tortoise coordinate)~\cite{Teukolsky:1973ha, Teukolsky:1974yv}. Thus, the appropriate boundary condition for the near-horizon solution to a BH spacetime 
demands $c_2$ be zero. Therefore, the radial function consistent with the horizon boundary condition, reduces the solution presented in \ref{radialfngen_1} to
\begin{equation}
\,_{s}R_{lm}(z) = c_1 z^{-s} (z+1)^{-N_3-s}  \, _2F_1\left(l-s-N_2+1,-l-s-N_1;-s+2 i P_++1;-z\right)~.
\end{equation}
To calculate the response of the Kerr BH to the external spin-$s$ tidal field, we first obtain the radial part of the field quantity $\zeta^{(s)}$ from the above solution and then determine the same in the intermediate region (this corresponds to large $r$ or, equivalently, large $z$ limit), which gives
\begin{multline}\label{zeta_gen}
\zeta_{\rm Intermediate}^{(s)} \propto z^{l-| s|+N_1 -N_3} \left\{\frac{\Gamma \left(-s+2 i P_++1\right) \Gamma \left(2 l+N_1-N_2+1\right)}{\Gamma \left(l+2 i P_++N_1+1\right) \Gamma \left(l-s-N_2+1\right)}\right\}\\\times\left[1+z^{-2l-1+N_2-N_1} \left\{\frac{ \Gamma \left(-2 l-N_1+N_2-1\right)\Gamma \left(l+2 i P_++N_1+1\right) \Gamma \left(l-s-N_2+1\right)}{\Gamma \left(-l-s-N_1\right) \Gamma \left(-l+2 i P_++N_2\right)\Gamma \left(2 l+N_1-N_2+1\right) }\right\}\right]\,.
\end{multline}
The above expansion for the radial part of the perturbation variable $\zeta^{(s)}$ has the term $z^{-2l-1}$, which also involves the linear-in-frequency and logarithmic extra piece $(N_{2}-N_{1})\ln{z}$. 
Therefore, from the coefficient of the $z^{-2l-1}$ term, the tidal response function becomes
\begin{equation}\label{resp_func_arb_rot_log}
\,_{s}F^{\rm Kerr}_{lm} = \frac{ \Gamma \left(-2 l-N_1+N_2-1\right)\Gamma \left(l+2 i P_++N_1+1\right) \Gamma \left(l-s-N_2+1\right)}{\Gamma \left(-l-s-N_1\right) \Gamma \left(-l+2 i P_++N_2\right)\Gamma \left(2 l+N_1-N_2+1\right)}\left[1+(N_2-N_1)\ln{\left(\frac{r}{r_{+}-r_{-}}\right)}\right]\,.
\end{equation}
Hence, there are two terms in the tidal response function: one is log-independent, and another one is log-dependent. It is straightforward to verify that the above expression correctly reproduces the results of Ref.~\cite{Bhatt:2024yyz} 
for Schwarzschild and 
Kerr BHs for gravitational perturbations, where we had ignored the logarithmic piece following Ref.~\cite{Katagiri:2023umb}.

As advocated by the effective field theory literature, the gauge invariant tidal response function is identified with the coefficient of the logarithmic term in the dynamical tidal response function obtained from BH perturbation theory~\cite{Saketh:2023bul, Hui:2020xxx}. Since the logarithmic term is accompanied by the factor $(N_2-N_1)$, which is of $\mathcal{O}(M\omega)$, and because we are working in the linear-in-frequency regime, it follows that the logarithmic tidal response function can be written as
\begin{equation}
\,_{s}F^{\rm Kerr(log)}_{lm}=(N_2-N_1) \,_{s}F^{\rm static}_{lm}\ln{\left(\frac{r}{r_{+}-r_{-}}\right)}\,,
\end{equation}
where $\,_{s}F^{\rm static}_{lm}$ is the static limit of the tidal response function in \ref{resp_func_arb_rot_log} found by calculating its $\omega\to 0$ limit,
\begin{equation}\label{static_TLNs_1}
\,_{s}F^{\rm static}_{lm}=-(-1)^{s}i P_+\frac{ \Gamma \left(1+l+s\right)\Gamma \left(l-s+1\right)}{\Gamma \left(2 l+2\right) \Gamma \left(2 l+1\right) }\prod_{j=1}^{l} (j^2+4P_+^2)\,,
\end{equation}
where $P_{+}=am/(r_+-r_-)$. In addition, the combination $N_{2}-N_{1}$ is given by 
\begin{equation}
N_2 -N_1 = am\omega\,\frac{4 (l^2+l+s^2)}{l(l+1)(2l+1)} + i M\omega\,\frac{8 (s-1) \left(r_+-M\right)}{(2 l+1)M}
\end{equation}
Since $N_2 - N_1$ has both real and imaginary terms, while $\,_{s}F^{\rm static}_{lm}$ is purely imaginary, the above written tidal response function with logarithmic piece will have non-zero real part too (or non-vanishing TLNs), which is proportional to $M\omega$. Thus, the tidal Love numbers with the logarithmic piece can be written as
\begin{equation}\label{dynamic_log_TLN}
\,_{s}k_{lm}^{\rm (log)}=(-1)^{s}am\omega\,\ln{\left(\frac{r}{r_{+}-r_{-}}\right)}\,\frac{4(s-1)\left(r_+-M\right)}{(2 l+1)\left(r_+-r_-\right)}\frac{ \Gamma \left(1+l+s\right)\Gamma \left(l-s+1\right)}{\Gamma \left(2 l+2\right) \Gamma \left(2 l+1\right) }\prod_{j=1}^{l} \left\{j^2+4\left(\frac{am}{r_+-r_-}\right)^2\right\}\,.
\end{equation}
Note that the TLNs (with the logarithmic piece) vanish for Schwarzschild $(a=0)$ and slowly rotating Kerr BH. In addition, they also vanish for non-extremal Kerr BH in axisymmetric tidal environment $(m=0)$. Intriguingly, the above result shows that for $s=1$ the TLNs associated with the logarithmic piece vanish identically. However, they do not vanish for $s=-1$, showing the disparity between $\pm|s|$ perturbation modes. Thus, non-extremal Kerr black holes also have non-zero TLNs in electromagnetic tidal field. In addition, the result presented in the above expression is not valid for scalar case, as mentioned earlier. We will study the scalar case separately in \ref{tide_scalar}.

We will now present the part of the TLNs, which does not involve the logarithmic piece, and from \ref{resp_func_arb_rot_log}, such log-independent response function yields (see also Ref.~\cite{Katagiri:2023umb})
\begin{equation}\label{resp_func_arb_rot}
\,_{s}F^{\rm Kerr\,(non-log)}_{lm} = \frac{ \Gamma \left(-2 l-N_1+N_2-1\right)\Gamma \left(l+2 i P_++N_1+1\right) \Gamma \left(l-s-N_2+1\right)}{\Gamma \left(-l-s-N_1\right) \Gamma \left(-l+2 i P_++N_2\right)\Gamma \left(2 l+N_1-N_2+1\right)}\,.
\end{equation}
To analyze further the tidal response function, we note that it can only be trusted up to linear order in $M\omega$, and hence we expand the tidal response function, keeping only terms linear orders of $M\omega$, which yields (for further details, see \ref{app_A.1})
\begin{align}\label{response_1}
\,_{s}F^{\rm Kerr\,(non-log)}_{lm}&=(2i P_+)\alpha\kappa\frac{ \Gamma \left(1+l+s\right)\Gamma \left(l-s+1\right)}{\Gamma \left(2 l+2\right) \Gamma \left(2 l+1\right) }\prod_{j=1}^{l} (j^2+4P_+^2)\Big[1+N_1\psi(1 +l + s)+N_1\psi(1 + l+2 i P_+)
\nonumber
\\
&- N_2\psi(1 + l - s)-N_2\psi(1 + l-2 i P_+)- (N_1 - N_2)\psi(2 + 2 l)- (N_1 - N_2)\psi(1 + 2 l)\Big]\,,
\end{align}
where we have introduced the quantities $\alpha$ and $\kappa$, having the following expressions:
\begin{equation}\label{def_alpha_kappa}
\alpha=\frac{\sin\pi(l+s+ N_1)}{\sin\pi(1 + 2 l + N_1 - N_2)}\,; 
\qquad 
\kappa = \frac{\sin\pi(-l+2 i P_++N_2)}{\sin(2i\pi P_+)}\,. 
\end{equation}
Just like gravitational perturbations, for generic spin perturbations as well the static and the dynamical TLNs branch out at this point~\cite{Bhatt:2024yyz}. To see this clearly, consider first the static TLNs, obtained from the $\omega\to0$ limit of the tidal response function and then determining its real part. In this limit, we obtain, $P_{+}\to am/(r_{+}-r_{-})$, $\alpha\to(-1)^{1-l+s}(1/2)$, and $\kappa\to(-1)^{l}-\sin(\pi l)\coth(2i\pi P_{+})$. So far, we had taken $l$ to be a complex quantity, but now with all the divergences canceled out, we can take $l\in \mathbb{Z}^{+}$, and $s\in\mathbb{Z}$, while satisfying $l\ge |s|$, such that the static tidal response function becomes the one presented in \ref{static_TLNs_1}. 

In the dynamical scenario ($\omega\ne0$), considering $M\omega$ to be small, we can write down $\alpha$ and $\kappa$ as
\begin{equation}\label{alkap}
\alpha=-(-1)^{-l+s}\frac{N_1}{N_1 - N_2}\,; 
\qquad 
\kappa=(-1)^l\left[1-i\pi N_2\coth(2 P_+\pi)\right]\,.
\end{equation}
To arrive at the above expression, we have considered $l\in \mathbb{Z}^{+}$, $s\in\mathbb{Z}$, and $l\ge |s|$, since all the divergent pieces have been removed. Thus, in the dynamical case, we have used the result $\omega\neq 0$ as well as $l\in \mathbb{Z}^{+}$, to obtain the form of $\alpha$ presented above. The static limit of this expression will be different if we had first taken $\omega \to 0$ limit and then $l\in \mathbb{Z}^{+}$, as in the static case. In short, we have the result 
\begin{equation}\label{limitissue}
\lim_{l\to \mathbb{Z}^{+}}\lim_{\omega \to 0}\alpha\neq \lim_{\omega \to 0} \lim_{l\to \mathbb{Z}^{+}} \alpha\,,
\end{equation}
implying that the static limit does not commute with integer $l$ limit. The same applies to TLNs as well, as we will demonstrate, zero frequency TLNs with integer $l$ need not match with the zero frequency limit of dynamical TLNs with integer $l$. This non-commuting limit is a feature of dynamical TLNs, arising from analytic continuation~\cite{Pani:2018inf, Chakraborty:2023zed}.

As evident from \ref{static_TLNs_1}, the static tidal response function is purely imaginary, and, hence, the static TLNs, which correspond to the real part of the static tidal response function, will turn out to be zero. Thus, the {\em static} TLNs of an arbitrarily rotating BH are vanishing for all possible external tidal perturbations. Thus, the vanishing of static TLNs for generic Kerr BH holds for scalar, electromagnetic, and gravitational perturbations, and is a feature of BHs in general relativity. It is to be emphasized that the above result correctly reproduces the ones presented in Refs.~\cite{LeTiec:2020spy, LeTiec:2020bos}.

In the dynamical scenario, one would like to discuss three possible limits, the Schwarzschild limit, the slow-rotation limit and the axisymmetric limit. We will discuss each of these in detail in the next section, while here we present the general result for the dynamical tidal response function under generic spin perturbations. Using the forms of $\alpha$ and $\kappa$ from \ref{alkap}, we can rewrite the dynamical tidal response function of \ref{response_1}, which does not involve the logarithmic piece, as
\begin{multline}\label{response_New1_a}
\,_{s}F^{\rm Kerr\,(non-log)}_{lm} = -(-1)^{s}(2i P_+)\frac{N_1}{N_1 - N_2}\left[1-i\pi N_2\coth(2 P_+\pi)\right]\frac{ \Gamma \left(1+l+s\right)\Gamma \left(l-s+1\right)}{\Gamma \left(2 l+2\right) \Gamma \left(2 l+1\right) }\prod_{j=1}^{l} (j^2+4P_+^2)\\\times\left[1+N_1\psi(1 +l + s)+N_1\psi(1 + l+2 i P_+)- N_2\psi(1 + l - s)-N_2\psi(1 + l-2 i P_+)\right.\\\left.- (N_1 - N_2)\psi(2 + 2 l)- (N_1 - N_2)\psi(1 + 2 l)\right].
\end{multline}
To analyze the tidal response function further, we need to examine its real and imaginary parts, which are related to the tidal Love numbers and tidal dissipation numbers, respectively. We will discuss this in the next section. 

\subsection{Dynamical tidal deformation and dissipation}
\label{sec:Tidal_response_arb_rot_BH_cal_real_Imag}

Having calculated the dynamical tidal response function in the last section, we will analyze some specific cases of interest in this section, namely, the Schwarzschild BH $(a=0)$, slowly rotating Kerr BH $(a\ll M)$, BH in axisymmetric tidal field ($m=0$), besides the general case. As already emphasized in the previous section, we will consider $s\neq0$ in the present section, which will be taken up separately in the next section. Note that $P_+$ is a small quantity in all three cases: while it is of $\mathcal{O}(M\omega)$ for a Schwarzschild BH and a BH in an axisymmetric tidal field, it is of $\mathcal{O}(a/M)$ and $\mathcal{O}(M\omega)$ for a slowly rotating Kerr BH. Similarly, the quantities $N_1$ and $N_2$ are also of $\mathcal{O}(M\omega)$, and hence we can neglect the terms like $P_{+}N_{1}$ and $P_{+}N_{2}$, respectively. Thus, for all of these cases, the dynamical tidal response function without the logarithmic piece reduces to
\begin{align}
\,_{s}F^{\text{Sch, Kerr(slow, axi)}}_{lm\,\textrm{(non-log)}}&=-(-1)^{s}(2i P_+)\left(\frac{N_1}{N_1-N_2}\right)\Big[1-i\pi N_2\coth(2 P_+\pi)\Big]
\nonumber
\\
&\qquad \qquad \times \frac{ \Gamma \left(1+l+s\right)\Gamma \left(l-s+1\right)}{\Gamma \left(2 l+2\right) \Gamma \left(2 l+1\right) }\prod_{j=1}^{l}(j^2)\,.
\end{align}
To proceed further, we express the term $\coth (2\pi P_+)$ as
\begin{equation}\label{cothx}
\coth (2\pi P_+) = \frac{1}{2\pi P_+} + \frac{1}{\pi}\sum_{k=1}^{\infty}\left(\frac{4P_+}{k^2 +4(P_+)^2}\right)\,,
\end{equation}
which follows from the fact that in all of the three cases $P_{+}$ is a small quantity. Substituting the above expression, the dynamical tidal response function for these three cases can be collectively expressed as
\begin{equation}
\,_{s}F^{\text{Sch, Kerr(slow, axi)}}_{lm\,\textrm{(non-log)}}=-(-1)^{s}\left(\frac{N_1}{N_1 - N_2}\right)\left(2iP_{+}+N_2\right)\frac{ \Gamma \left(1+l+s\right)\Gamma \left(l-s+1\right) (\Gamma(l+1))^2}{\Gamma \left(2 l+2\right) \Gamma \left(2 l+1\right) }\,,
\end{equation}
where we have neglected the second- and higher-order terms of $M\omega$. For all of these three cases, it follows that both $N_1$ and $N_2$ are purely imaginary, which in turn ensures that the dynamical tidal response function in these three limiting cases are also purely imaginary in nature. Implying vanishing of the dynamic TLNs for all of these three cases,
\begin{equation}
\,_{s}k^{\text{Sch, Kerr(slow, axi)}}_{lm\,\textrm{(non-log)}}=\frac{1}{2}\textrm{Re}\,\,_{s}F^{\text{Sch, Kerr(slow, axi)}}_{lm\,\textrm{(non-log)}}=0\,.
\end{equation}
We would like to reiterate that these results have been derived assuming non-zero spin for the tidal field. The scalar case will be discussed shortly. We would also like to point out that the above expression diverges for a slowly rotating black hole in electromagnetic tidal field with spin weight $s=1$. A separate analysis of the tidal response function [\ref{resp_func_arb_rot}] for this scenario will lead us to vanishing tidal Love numbers.

After the specific cases have been discussed in detail, we turn our attention to the determination of the TLNs for generic Kerr BH. Since these limits are singular for the scalar perturbation, the generic result will also be for the $s\neq 0$ case. To extract the real and imaginary parts of the dynamical tidal response function, we rewrite \ref{response_New1_a} as
\begin{equation}
\,_{s}F^{\textrm{(non-log)}}_{lm}=-(-1)^s(2iP_{+})\left(\frac{N_1}{N_1-N_2}\right)\Big[1-i\pi N_2\coth(2 P_+\pi)\Big]F_2\left(1+M\omega F_1\right)\,,
\end{equation}
where we have introduced two quantities $F_{1}$ and $F_{2}$, taking the form
\begin{align}\label{def_Mo_F1}
M\omega F_{1}\equiv N_1\psi(1 +l + s)&+N_1\psi(1 + l+2 i P_+)- N_2\psi(1 + l - s)-N_2\psi(1 + l-2 i P_+)\nonumber\\
&- (N_1 - N_2)\psi(2 + 2 l)- (N_1 - N_2)\psi(1 + 2 l)\,,
\end{align}
and
\begin{equation}\label{def_F2}
F_{2}\equiv \frac{ \Gamma \left(1+l+s\right)\Gamma \left(l-s+1\right)}{\Gamma \left(2 l+2\right) \Gamma \left(2 l+1\right) }\prod_{j=1}^{l} (j^2+4P_+^2)\,. 
\end{equation}
Using the fact that $N_1/(N_1-N_2)$, $F_{1}$ and $F_{2}$ are independent of the frequency, along with the result that $F_1$ is a complex quantity while $F_2$ is real, one can write down the dynamical TLNs of a non-extremal Kerr BH as
\begin{align}\label{eq_39_a}
\,_{s}k_{lm}^{\textrm{(non-log)}}&=\frac{1}{2}\textrm{Re} \,_{s}F_{lm}^{\textrm{(non-log)}}=-(-1)^s P_{+}F_{2}\,\text{Im}\Big[\left(\frac{N_1}{N_1-N_2}\right)\Big\{1-i\pi N_2\coth(2 P_+\pi)\Big\}\left(1+M\omega F_1\right)\Big]
\nonumber
\\
&=-(-1)^s P_{+}F_{2}\Bigg[-\pi\coth(2 P_+\pi)\text{Re}\left(\frac{N_1}{N_1-N_2}\right)\text{Re}N_{2}
+\text{Im}\left(\frac{N_1}{N_1-N_2}\right)\left\{1+\pi \coth(2 P_+\pi)\textrm{Im}N_{2}\right\}
\nonumber
\\
&+\left\{\text{Re}\left(\frac{N_1}{N_1-N_2}\right)\left\{1+\pi \coth(2 P_+\pi)\textrm{Im}N_{2}\right\}+\pi\coth(2 P_+\pi)\text{Im}\left(\frac{N_1}{N_1-N_2}\right)\text{Re}N_{2}\right\}\text{Im}(M\omega F_1)
\nonumber
\\
&+\left\{-\pi\coth(2 P_+\pi)\text{Re}\left(\frac{N_1}{N_1-N_2}\right)\text{Re}N_{2}+\text{Im}\left(\frac{N_1}{N_1-N_2}\right)\left\{1+\pi \coth(2 P_+\pi)\textrm{Im}N_{2}\right\}\right\}\text{Re}(M\omega F_1)\Bigg]~.
\end{align}
On further analyzing the above expression, and by removing second- and higher-order terms in the frequency, we can write down the dynamical TLNs of a non-extremal Kerr BH as (see \ref{app_A.2} for details)
\begin{equation}\label{dynamic_TLNs}
\,_{s}k_{lm}^{\textrm{(non-log)}} = (am)^2\,_{s}k^{(0)}_{lm} + am\omega\,_{s}k^{(1)}_{lm} + \mathcal{O}(M^{2}\omega^{2})~,
\end{equation}
where detailed expressions for the quantities $\,_{s}k^{(0)}_{lm}$, $\,_{s}k^{(1)}_{lm}$ have been provided in \ref{app_A.2}, and both are independent of the frequency scale $\omega$. The above given form of the dynamical TLNs correctly reproduce the corresponding expressions for the specific cases discussed earlier. However, one cannot obtain the static TLNs by taking the $\omega\to0$ limit of the above expression, since the above result have been derived strictly for non-zero frequencies [see \ref{limitissue} and the discussion thereof, for further details].

\begin{figure}
    \centering
    \begin{subfigure}[b]{0.475\textwidth}
        \centering
        \includegraphics[scale=0.7]{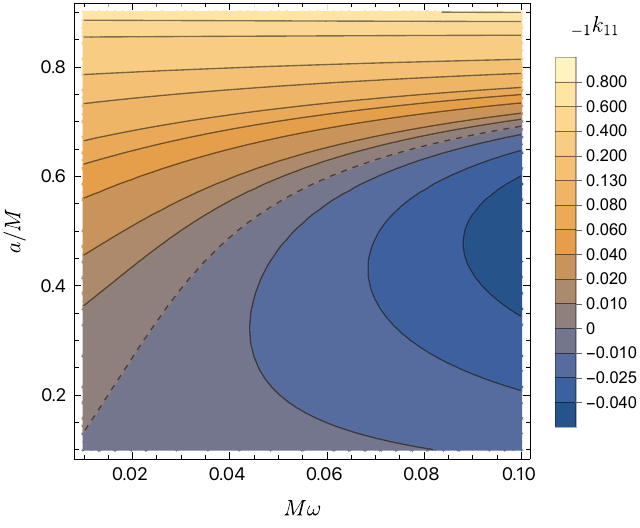}
        \caption{For spin-weight $-1$}
        \label{fig:1a}
    \end{subfigure}
    \hfill
    \begin{subfigure}[b]{0.475\textwidth}
        \centering
        \includegraphics[scale=0.7]{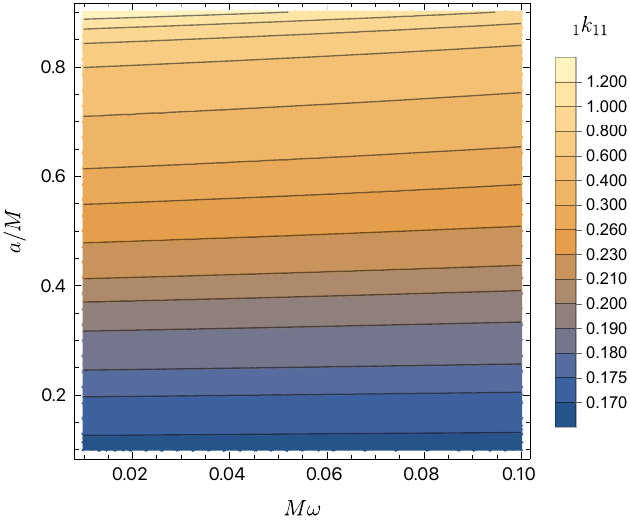}
        \caption{For spin-weight $1$}
        \label{fig:1b}
    \end{subfigure}
    \vskip\baselineskip
    \begin{subfigure}[b]{0.475\textwidth}
        \centering
        \includegraphics[scale=0.7]{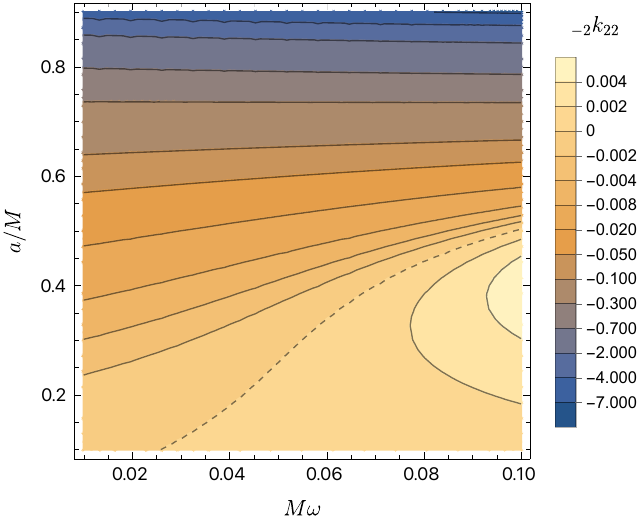}
        \caption{For spin-weight $-2$}
        \label{fig:1c}
    \end{subfigure}
    \hfill
    \begin{subfigure}[b]{0.475\textwidth}
        \centering
        \includegraphics[scale=0.7]{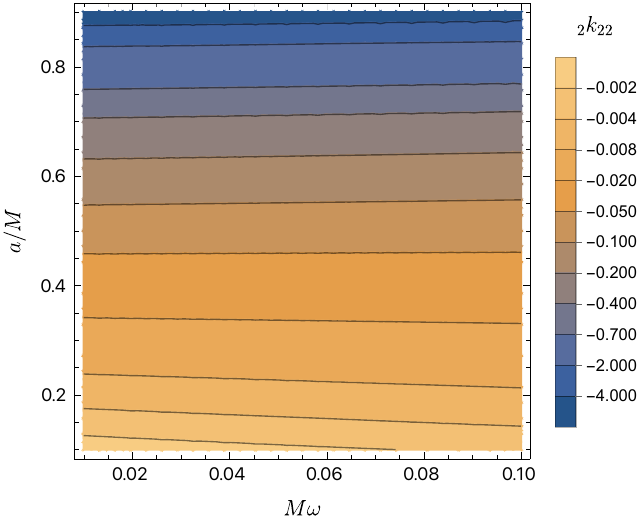}
        \caption{For spin-weight $2$}
        \label{fig:1d}
    \end{subfigure}
    \caption{Contour plots of TLNs (with non-logarithmic term) associated with a non-extremal Kerr BH have been presented with dimensionless frequency $M\omega$ and dimensionless rotation parameter $(a/M)$, for EM and gravitational perturbations with different spin-weights $s$. In the case of negative spin weights, for both EM and gravitational perturbations, TLNs can be zero, positive, or negative for some particular values of the frequency and rotation parameter. The zero TLN contours are shown with dashed lines in \ref{fig:1a} and \ref{fig:1c}. Note that for EM perturbation we have taken $l=1=m$, while for gravitational perturbation, we have $l=2=m$. We would also like to point out that our analysis is only valid for small mode frequencies, satisfying $M\omega \ll 1$. Keeping this in mind, we have presented the TLNs in the above plots till $M\omega\sim 0.1$. However, even for $M\omega \sim \mathcal{O}(0.1)$, the mode frequency is small but may not still be considered as a very small quantity. Thus, the features present here may also be influenced by non-linear terms, which we have ignored in our analysis.}
    \label{fig:1}
\end{figure}

One intriguing fact arising out of our analysis connects with the fact that the form of $\,_{s}k^{(0)}_{lm}$ and $\,_{s}k^{(1)}_{lm}$ changes its value if we change $s\to-s$. Thus, different spin perturbations affect the tidal response function differently. The implications can be far reaching; e.g., the tidal field described by $\Psi_{4}$ has different tidal response function compared to $\Psi_{0}$. It remains to be seen if this has anything to do with the axial and polar decomposition of the TLNs in the non-rotating limit. This feature can also be seen from \ref{fig:1}. As evident, the TLNs associated with perturbations having positive spin weights differ from those with negative spin weights. Moreover, for negative spin weights, for both EM and gravitational perturbations, the TLNs can be positive or negative [see \ref{fig:1a} and \ref{fig:1c}]. Intriguingly, in this case of negative spin weights, for small rotation and larger frequencies EM TLNs are negative, while for large rotation and small frequencies, they are positive, which is exactly opposite of what gravitational TLNs depict. On the other hand, for positive spin weights, TLNs associated with EM perturbations are positive, but the gravitational perturbations yield negative TLNs [see \ref{fig:1b} and \ref{fig:1d}]. We can also write the tidal Love numbers as the summation of the logarithmic term, presented in \ref{dynamic_log_TLN}, and the non-logarithmic term, as in \ref{dynamic_TLNs}, where the logarithmic term depends on $am\omega$, though these terms have been obtained using different limits for expanding the tidal response function.

In addition, the above computations have been performed for an non-extremal Kerr BH and hence can not be applied for an extremal Kerr BH in a straightforward manner. This is because of the choice of the coordinate $z$, which becomes ill-behaved in the extremal limit. Moreover, as already emphasized earlier, the above results hold for non-zero spin of the perturbation. Thus, we will study the tidal effects due to an external scalar perturbation, as well as for an extremal Kerr BH, separately in the next sections.

\section{Love numbers of rotating black hole to scalar tidal field}\label{tide_scalar}

In this section, we will study the response of a non-extremal Kerr BH to an external scalar tidal field. For scalar tidal field, with spin weight $s=0$, the radial Teukolsky equation, as presented in \ref{TZ}, becomes
\begin{equation}\label{TZ_0}
\frac{\mathrm{d}^2\,_{0}R_{lm}}{\mathrm{d}z^2} + \left[\frac{2iP_++1}{z}-\frac{2iP_--1}{1+z}-2i\omega(r_+-r_-)\right]\frac{\mathrm{d}\,_{0}R_{lm}}{\mathrm{d}z} +\left[\frac{A_-}{1+z}-\frac{A_+}{z}\right]\,_{0}R_{lm}=\frac{\,_{0}T_{lm}}{\Delta}(r_+-r_-)^2\,.
\end{equation}
For small mode frequency $(M\omega\ll1)$ and near-zone $(M\omega z\ll1)$ approximation, we can simplify the above equation to the form
\begin{equation}\label{TZ_0_1}
\frac{\mathrm{d}^2\,_{0}R_{lm}}{\mathrm{d}z^2} + \left[\frac{2iP_++1}{z}-\frac{2iP_1-1}{1+z}\right]\frac{\mathrm{d}\,_{0}R_{lm}}{\mathrm{d}z} -\frac{A_+}{z(1+z)}\,_{0}R_{lm}=0,
\end{equation}
where we have introduced the frequency dependent quantity $P_{1}\equiv P_{-}+ \omega(r_+-r_-)$, where $P_{-}$ has been defined earlier, and we have put the source term to zero.

Similar to the generic radial equation, presented in the last section, the above equation also has three regular singular points, located at $z = 0,~-1$, and $\infty$. Therefore, the solution of the above equation can be expressed in terms of the Hypergeometric function,
\begin{multline}
\,_{0}R_{lm}(z) = C_1 \,_{2}F_{1}\left[-l-U_1, 1+l-U_2; 1+2iP_+; -z\right] 
\\
+ C_2\,z^{-2iP_+}\,_{2}F_{1}\left[-l-2iP_+ - U_1, 1+l-2iP_+ -U_2; 1-2iP_+; -z\right]\,,
\end{multline}
where $C_1$ and $C_2$ are the constant of integration and the quantities $U_1$ and $U_2$ read
\begin{equation}\label{def_U1_and_U_2}
U_1 = 2ir_+\omega - \frac{2am\omega}{2l+1}\,;
\qquad 
U_2 = 2ir_+\omega + \frac{2am\omega}{2l+1}\,.
\end{equation}
Note that, unlike $N_{1}$ and $N_{2}$ in the previous section, the quantities $U_{1}$ and $U_{2}$ are well-behaved in the Schwarzschild and axisymmetric limit. 
Similar to the previous section, applying ingoing boundary condition at the horizon, which demands $C_2 = 0$, the above solution becomes
\begin{equation}
\,_{0}R_{lm}(z)=C_1 \,_{2}F_{1}\left[-l-U_1, 1+l-U_2; 1+2iP_+; -z\right]\,.
\end{equation}
From the large $r$ (or, large $z$) limit of the above solution, we can identify the tidal response function as the coefficient of $z^{-2l-1}$, which reads 
\begin{equation}\label{resp_func_arb_rot_s_0_log}
\,_{0}F_{lm} = \frac{\Gamma (-1-2l+U_2-U_1)\Gamma (1+l-U_2) \Gamma (1+l+2iP_++U_1)}{\Gamma (-l-U_1) \Gamma (-l+2iP_++U_2)\Gamma (1+2l+U_1-U_2)}\left[1+(U_2-U_1)\ln{\left(\frac{r}{r_{+}-r_{-}}\right)}\right]\,.
\end{equation}
To calculate the static tidal response function, we take the $\omega\to0$ limit of the above expression, which yields
\begin{equation}\label{static_resp_func_arb_rot_s_0}
\,_{0}F_{lm}^{\rm static} = \frac{\Gamma (-1-2l)\Gamma (1+l) \Gamma (1+l+2iP_+)}{\Gamma (-l) \Gamma (-l+2iP_+)\Gamma (1+2l)} \equiv -i P_+\frac{ \Gamma (1+l)\Gamma (l+1)}{\Gamma (1 + 2l) \Gamma (2l+2)} \prod_{j=1}^{l} (j^2+4P_+^2)\,,
\end{equation}
with $P_+ = am/(r_+-r_-)$. Since the static tidal response function is purely imaginary, the TLNs of non-extremal Kerr black holes vanish for a static scalar tidal field. 

The logarithmic contribution to the tidal response function, given the above result, is then given by $(U_2-U_1)\,_{0}F_{lm}^{\rm static} \ln{\{r/(r_+-r_-)\}}$. Since $U_2-U_1$ is a real quantity and $\,_{0}F_{lm}^{\rm static}$ is purely imaginary, we will not have any logarithmic contribution to the tidal Love numbers. Now, we will study the non-logarithmic part of the tidal response function, i.e.,
\begin{equation}\label{resp_func_arb_rot_s_0}
\,_{0}F_{lm} = \frac{\Gamma (-1-2l+U_2-U_1)\Gamma (1+l-U_2) \Gamma (1+l+2iP_++U_1)}{\Gamma (-l-U_1) \Gamma (-l+2iP_++U_2)\Gamma (1+2l+U_1-U_2)}\,.
\end{equation}
As in the case of EM and gravitational perturbations, here, too, we study specific cases of \ref{resp_func_arb_rot_s_0} for the dynamical scenario $(\omega \ne 0)$. For Schwarzschild BH $(a = 0)$, it follows that $U_1 = 4iM\omega = U_2$, and $P_+ = -2M\omega$. 
Thus for the Schwarzschild BH, the above equation becomes
\begin{equation}\label{resp_func_arb_rot_s_0_Schw}
\,_{0}F_{lm}^{\rm{Schw}} = \frac{\Gamma (-1-2l)\Gamma (1+l+2iP_+) \Gamma (1+l)}{\Gamma (-l+2iP_+) \Gamma (-l)\Gamma (1+2l)} \equiv -i P_+\frac{ \Gamma (1+l)\Gamma (l+1)}{\Gamma (1 + 2l) \Gamma (2l+2)} \prod_{j=1}^{l} (j^2+4P_+^2)\,.
\end{equation}
Therefore, the scalar tidal response function for Schwarzschild BH is purely imaginary in nature, implying vanishing scalar TLNs. 

On the other hand, for axisymmetric tidal perturbation $(m=0)$, we obtain the results $U_{1}=2ir_{+}\omega=U_{2}\equiv U$ and $P_{+}=-2Mr_{+}\omega/(r_+-r_-)$, up to linear order in frequency. Similarly, for slowly rotating Kerr BH, $P_+ = (am-4M^2\omega)/2M$ and $U_1 \sim 4iM\omega \sim U_2\equiv U$, where we have ignored the second- and higher-order terms of $\mathcal{O}(M\omega)$. Thus, the tidal response function for the slowly rotating Kerr BH and non-extremal Kerr BH in an axisymmetric tidal field becomes
\begin{equation}\label{resp_func_arb_rot_s_0_axi}
\,_{0}F_{lm}^{\rm{Kerr(slow),~Kerr(axi)}}=\frac{\Gamma (-1-2l)\Gamma (1+l-U) \Gamma (1+l+2iP_++U)}{\Gamma (-l-U) \Gamma (-l+2iP_{+}+U)\Gamma (1+2l)}\sim -i P_+\frac{ (\Gamma (1+l))^4}{\Gamma (1 + 2l) \Gamma (2l+2)}\,,
\end{equation}
where we have expanded the gamma functions containing $\mathcal{O}(M\omega)$ terms and have neglected the second- and higher-order terms of $\mathcal{O}(M\omega)$.
As evident, the above tidal response function is also purely imaginary in nature, i.e., the scalar TLNs of a Kerr BH under axisymmetric tidal perturbation identically vanish. It also implies that the scalar TLNs of slowly rotating Kerr BH vanish as well. Thus, it is interesting to note that the behavior of the scalar tidal response function for a Kerr black hole in the non-rotating, slowly rotating, and axisymmetric tidal field limits yields zero TLNs, as in the case of perturbations with non-zero spins. Hence, we can conclude that in these three limiting cases the TLNs of a non-extremal Kerr BH identically vanish, irrespective of the spin of the perturbation. 

Returning to the case of generic, non-axisymmetric tidal perturbation, we can expand the gamma functions in \ref{resp_func_arb_rot_s_0} to the linear order of $M\omega$ and hence obtain the expression for the scalar tidal response function
\begin{equation}
\,_{0}F_{lm}=(-2iP_{+})\left(\frac{U_1}{U_1-U_2}\right)\Big[1-i\pi U_2\coth(2 P_+\pi)\Big]F_2\left(1+M\omega F_1\right)\,,
\end{equation}
where the ratio $\{U_{1}/(U_{1}-U_{2})\}$ is independent of $\omega$ and the quantities $F_{1}$ and $F_{2}$ are given by
\begin{align}\label{def_Mo_F1_s_0}
M\omega F_{1}\equiv U_1\psi(1 +l)&+U_1\psi(1 + l+2 i P_+)
- U_2\psi(1 + l)-U_2\psi(1 + l-2 i P_+)
\nonumber 
\\
&- (U_1 - U_2)\psi(2 + 2 l)- (U_1 - U_2)\psi(1 + 2 l)\,
\end{align}
and
\begin{equation}\label{def_F2_s_0}
F_{2}\equiv \frac{(l!)^2}{\left(2 l+1\right)!\left(2l\right)!}\prod_{j=1}^{l} (j^2+4P_+^2)\,. 
\end{equation}
As worked out in the last section, here also we can calculate the TLNs from the real part of the tidal response function, which reads
\begin{equation}\label{TLNs_non_extremal_Kerr_scalar_non_axi_tidal_field_0}
\,_{0}k_{lm}=\frac{1}{2}\textrm{Re} \,_{0}F_{lm}=-P_{+}F_{2}\,\text{Im}\Big[\left(\frac{U_1}{U_1-U_2}\right)\Big\{1-i\pi U_2\coth(2 P_+\pi)\Big\}\left(1+M\omega F_1\right)\Big]~.
\end{equation}
On further simplification, we can express it as
\begin{equation}\label{TLNs_non_extremal_Kerr_scalar_non_axi_tidal_field}
\,_{0}k_{lm} = \,_{0}k_{lm}^{(0)} + M\omega \,_{0}k_{lm}^{(1)} + \mathcal{O}(M^2\omega^2)\,,
\end{equation}
where the explicit expressions of $\,_{0}k_{lm}^{(0)}$ and $\,_{0}k_{lm}^{(1)}$ are given in \ref{appendix_B_New}. We would like to emphasize that the above equation is only valid for non-zero mode frequencies ($\omega$) and non-zero azimuthal indices ($m$). Hence, one cannot obtain the $\omega\to0$ and $m\to0$ limit from the above equation. Note that we have already calculated the tidal response functions for the static case $(\omega =0 )$ and the axisymmetric tidal field $(m=0)$ in \ref{static_resp_func_arb_rot_s_0} and \ref{resp_func_arb_rot_s_0_axi}, respectively. The above result implies that the TLNs are non-vanishing for rotating BHs living in the non-axisymmetric scalar tidal field. Thus, in general, for generic tidal perturbation, the TLNs of a rotating non-extremal BH are non-zero. 
\begin{figure}[ht]
    \centering
    \begin{subfigure}[b]{0.475\textwidth}
        \centering
        \includegraphics[scale=0.7]{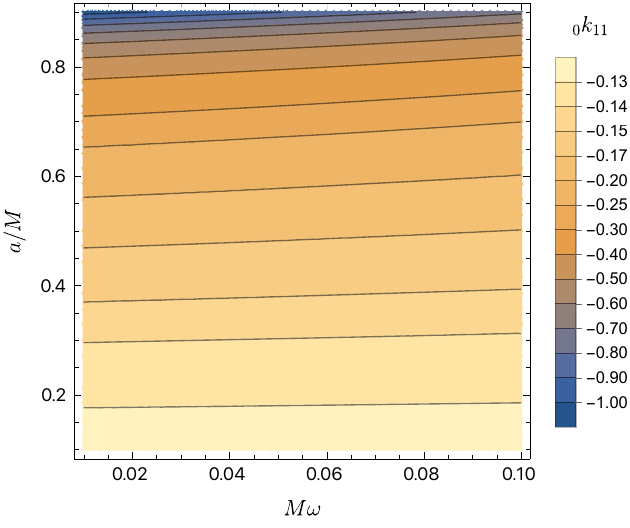}
        \caption{For $l=1=m$}
        \label{fig:2a}
    \end{subfigure}
    \hfill
    \begin{subfigure}[b]{0.475\textwidth}
        \centering
        \includegraphics[scale=0.7]{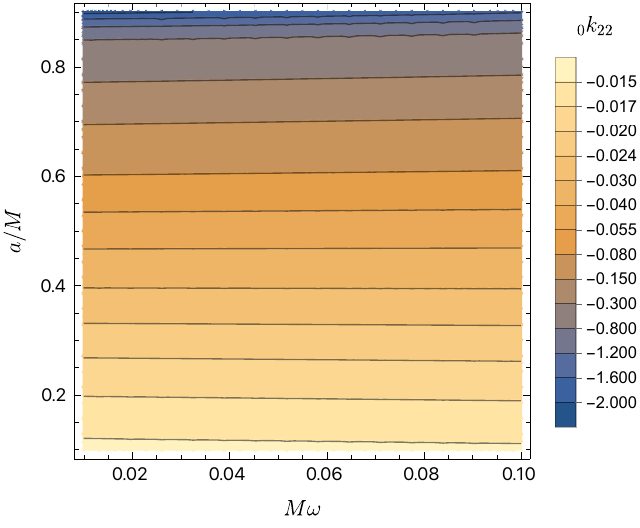}
        \caption{For $l=2=m$}
        \label{fig:2b}
    \end{subfigure}
    \caption{Contour plots of the TLNs $\,_{0}k_{11}$ (associated with the $l=1=m$ mode of the scalar perturbation) and $\,_{0}k_{22}$ (associated with the $l=2=m$ mode of the scalar perturbation) are presented against the dimensionless frequency $M\omega$ and the dimensionless rotation parameter $(a/M)$. As is evident, the TLNs are negative, and their magnitudes increase with increasing rotation. Here, too, we have plotted the TLNs for mode frequencies up to $M\omega\sim 0.1$, which, though small, may not be considered as very small compared to unity. Thus, the behavior of the TLNs at the rightmost region in these plots may be affected by non-linear terms we have ignored in our analysis.}
    \label{fig_2}
\end{figure}

There is one interesting aspect, though, associated with the scalar TLNs of Kerr BH. Namely, in this case, the fundamental mode ($l=0=m$) has vanishing TLN. The non-zero TLNs for scalar perturbation can be found for higher modes, e.g., the $l=1=m$ mode. This has also been presented in \ref{fig_2}. Intriguingly, the Love number associated with the $l=1=m$ mode is negative, and its magnitude increases as the rotation parameter of the BH also increases. Similar behavior is also shown by the Love number associated with the $l=2=m$ mode.

\section{Dynamical Love numbers of an extremal Kerr black hole}\label{sec:Tidal_response_ext_rot_BH_cal}

As discussed previously, the new coordinate $z$ defined in the last section ($z\equiv(r-r_+)/(r_+-r_-)$) is not well-behaved for an extremal Kerr BH because it diverges for $r_{+}=r_{-}$. Therefore, we need to study the particular case of extremal Kerr BH separately, which we accomplish in this section. To have a well-behaved coordinate system, we introduce the following coordinate: $\bar{z}\equiv (r-r_+)/r_+$. 
In addition, for an extremal Kerr BH, the quantity $\Delta$ becomes $\Delta=(r-r_{+})^{2}$, where $r_{+}=r_{-}=M$. Thus, for an extremal Kerr BH, the source free radial Teukolsky equation, as presented in \ref{TEq}, reads
\begin{multline}\label{ext_BH_eq}
\frac{\mathrm{d}^2\,_{s}R_{lm}}{\mathrm{d}\bar{z}^2} + \left[\frac{2\{(s+1)-2iM\omega\}}{\bar{z}} + \frac{2i(m-2M\omega)}{\bar{z}^2} - 2iM\omega\right]\frac{\mathrm{d}\,_{s}R_{lm}}{\mathrm{d}\bar{z}} \\+\left[\frac{-2 iM\omega}{\bar{z}}+\frac{-\lambda -2 i M (2 s+1) \omega}{\bar{z}^2}+\frac{4 i s (m-2 M \omega )}{\bar{z}^3}\right]\,_{s}R_{lm}=0~.
\end{multline}
To proceed further, here also we can use the following approximations: (a) small frequency $(M\omega\ll1)$ and (b) near-zone $(M\omega \bar{z}\ll1)$, such that the above radial equation reduces to
\begin{multline}\label{approx_ext_BH}
    \frac{\mathrm{d}^2\,_{s}R_{lm}}{\mathrm{d}\bar{z}^2} + \left[\frac{2\{(s+1)-2iM\omega\}}{\bar{z}} + \frac{2i(m-2M\omega)}{\bar{z}^2}\right]\frac{\mathrm{d}\,_{s}R_{lm}}{\mathrm{d}\bar{z}} \\+\left[\frac{-\lambda -2 i M (2 s+1) \omega}{\bar{z}^2}+\frac{4 i s (m-2 M \omega )}{\bar{z}^3}\right]\,_{s}R_{lm} = 0\,,
\end{multline}
where the separation constant $\lambda$ between the radial and the angular Teukolsky equation becomes
\begin{equation}\label{def_lambda_1}
\lambda = l (l+1)-s (s+1)-2 m M \omega  \left\{\frac{s^2}{l (l+1)}+1\right\}\,.
\end{equation}
To find out the tidal response function associated with generic spin perturbation of the extremal Kerr BH, we note that \ref{approx_ext_BH} is a second order differential equation with one regular singular point, located at $\bar{z}=\infty$, and an irregular singular point, at $\bar{z}=0$. Thus, the solution of the above differential equation can be expressed as a confluent hypergeometric function,
\begin{multline}
\,_{s}R_{lm}(\bar{z})=\bar{z}^{-\frac{1}{2}\left(1+2s-4iM\omega+\beta\right)}\left[c_1 U\left(\frac{1}{2}\left(1-2s-4iM\omega+\beta\right),1+ \beta, \frac{2i(m-2M\omega)}{\bar{z}}\right) \right.\\\left.+ c_2 e^{\frac{2i(m-2M\omega)}{\bar{z}}} U\left(\frac{1}{2}\left(1+2s+4iM\omega+\beta\right),1+ \beta,-\frac{2i(m-2M\omega)}{\bar{z}}\right)\right]~,
\end{multline}
where $c_1$ and $c_2$ are the constants of integration, and we have introduced two frequency dependent constants $\beta$ and $N$, which read
\begin{align}\label{def_beta_a}
\beta&\equiv
2l+1-2N+\mathcal{O}(M^{2}\omega^{2})~;
\quad
N\equiv \frac{2mM\omega\left(l^2+l+s^2\right)}{l(l+1)(2l+1)}~.
\end{align}
Since we are working with BHs, we can further simplify the solution by implementing the purely ingoing boundary condition at the horizon. 
To implement the same, we will require the $r\to r_{+}$ (or, $z\to 0$) limit of the above solution, which yields
\begin{equation}\label{R_near_ext}
\,_{s}R_{lm}^{\rm near}(\bar{z})\sim c_1 \, \mathcal{C}^{-\frac{1}{2}\left(1-2s-4iM\omega+\beta\right)} \bar{z}^{-2s} + c_2 e^{\mathcal{C}/\bar{z}} (-\mathcal{C})^{-\frac{1}{2}\left(1+2s+4iM\omega+\beta\right)}\bar{z}^{4iM\omega}~,
\end{equation}
where $\mathcal{C}\equiv 2i(m-2M\omega)$. The first term contains $\bar{z}^{-2s}$, which arises from $\Delta^{-s}$, and hence corresponds to the ingoing mode at the event horizon~\cite{Teukolsky:1974yv}. On the other hand, the outgoing mode at the horizon $\exp[-2i(m-2M\omega)(r_{*}/2M)]$, where $r_*$ is tortoise coordinate, behaves as $e^{\mathcal{C}/\bar{z}} \bar{z}^{4iM\omega}$ near the horizon.\footnote{Note that the tortoise coordinate can be expressed in terms of the radial coordinate $r$ as, 
\begin{equation}
r_{*}=r+2M \ln{\left(\frac{r-M}{M}\right)}-\frac{2M^{2}}{r-M}\simeq M+2M\ln{\bar{z}}-\frac{2M}{\bar{z}}\,.
\end{equation}
Therefore, we can write down the outgoing term as $e^{-2i(m-2M\omega)(r_{*}/2M)}\sim  e^{\mathcal{C}/\bar{z}}\bar{z}^{4iM\omega}\bar{z}^{-2im}$. Note that the term $\bar{z}^{-2im}$ is unaccounted for and is a general feature of extremal BHs.} Thus, the second term in \ref{R_near_ext} is the outgoing mode at the event horizon, implying that $c_2$ must be zero, which brings us to the following solution of the radial Teukolsky equation for generic perturbation of Kerr BH,
\begin{equation}\label{solution_final_ext_BH}
\,_{s}R_{lm}(\bar{z})=c_1 \bar{z}^{-\frac{1}{2}\left(1+2s-4iM\omega+\beta\right)} U\left(\frac{1}{2}\left(1-2s-4iM\omega+\beta\right),1+ \beta, \frac{\mathcal{C}}{\bar{z}}\right)~.
\end{equation}
The subsequent computation of the tidal response function for an extremal Kerr BH proceeds identically, where one determines the coefficient of $\bar{z}^{-\beta}$ in the intermediate region (large $r$ or large $\bar{z}$ limit). Here, too, in the linear order of $M\omega$, $\bar{z}^{-\beta}$ can be written as $\bar{z}^{-2l-1}(1+2N\ln{\bar{z}})$, which will yield the following tidal response function:
\begin{align}\label{non_static_rf_ext_BH_log}
\,_{s}F_{lm}&=-\{2i(m-2M\omega)\}^{2l+1-2N}\frac{\Gamma \left(-2l+2N\right) \Gamma \left(1+l-s-2 i M \omega-N\right)}{\Gamma \left(-l-s-2 i M \omega+N\right)\Gamma \left(2l+2 -2N\right)}\left[1+2N\ln{\left(\frac{r}{r_{+}}\right)}\right]\,.
\end{align}
The static tidal response function can be determined by taking the $\omega\to0$ limit of the above expression, which yields (keeping in mind that $N$ is linearly proportional to $\omega$)
\begin{equation}
\,_{s}F_{lm}^{\rm static}=-(-1)^{s-l}im(2im)^{2l}\frac{\Gamma (l-s +1) \Gamma (1+l+s)}{\Gamma (2l+1)\Gamma (2l+2)}\,.
\end{equation}
It is apparent that the above static tidal response function is purely imaginary in nature, which implies vanishing static tidal Love numbers for extremal Kerr BHs under generic perturbations. 

The logarithmic part of the tidal response function, on the other hand, is given by $2N\,_{s}F_{lm}^{\rm static} \ln{(r/r_+)}$. Since $N$ is a real quantity and $\,_{s}F_{lm}^{\rm static}$ is purely imaginary, we will not have any logarithmic contribution to the tidal Love numbers. Hence, the logarithmic term in the dynamical tidal Love numbers only appear for non-extremal Kerr black holes in electromagnetic $(s=-1)$ and gravitational $(s=\pm2)$ tidal fields. Now, we will study the non-logarithmic part of the tidal response function, i.e.,
\begin{align}\label{non_static_rf_ext_BH}
\,_{s}F_{lm}&=-\{2i(m-2M\omega)\}^{2l+1-2N}\frac{\Gamma \left(-2l+2N\right) \Gamma \left(1+l-s-2 i M \omega-N\right)}{\Gamma \left(-l-s-2 i M \omega+N\right)\Gamma \left(2l+2 -2N\right)}\,.
\end{align}
Similar to the case of non-extremal Kerr BHs, the dynamical case  ($M\omega\ne0$) needs to be studied separately. To start with, we first expand the tidal response function in the linear order of $M\omega$ (for intermediate steps, see \ref{app_B.1}), yielding
\begin{multline}\label{simplified_non_static_rf_ext_BH}
\,_{s}F_{lm}=(-1)^{l+s}i^{2l+1-2N}\left\{2(m-2M\omega)\right\}^{2l+1-2N}\left(\frac{\sin[(2iM\omega-N)\pi]}{\sin[(2N)\pi]}\right)\frac{\Gamma \left(1+l+s\right) \Gamma \left(1+l-s\right)}{\Gamma \left(1+2l\right)\Gamma \left(2l+2\right)}
\\
\times\left[1+(2 i M \omega-N)\psi(1+l+s)+(-2 i M \omega-N)\psi(1+l-s)+2N\psi(1+2l)+2N\psi(2+2l)\right]\,.
\end{multline}
Since $M\omega$ is very small, and for strictly dynamical case, where $M\omega \ne 0$, we obtain 
\begin{equation}\label{def_alpha_1b}
\frac{\sin[(2iM\omega-N)\pi]}{\sin[(2N)\pi]}\simeq \frac{2iM\omega-N}{2N}
=\frac{1}{2}\left[-1+i\frac{l(l+1)(2l+1)}{m\left(l^2+l+s^2\right)}\right]\,.
\end{equation}
Therefore, having gotten rid of any divergent terms, we may use the results $l\in \mathbb{Z}^{+}$, $s\in\mathbb{Z}$, and $l\ge |s|$ and hence write $i^{2l+1-2N}=(-1)^{l}ie^{-iN\pi}\approx(-1)^{l}i\left[1-iN\pi\right]$, for small $M\omega$, using which we can re-express the tidal response function as
\begin{multline}\label{simplified_non_static_rf_ext_BH_02}
\,_{s}F_{lm}=-(-1)^s i\left(1-iN\pi\right)(2m)^{2l-2N}\left\{m-(2l+1)2M\omega\right\}\left[1-i\frac{l(l+1)(2l+1)}{m\left(l^2+l+s^2\right)}\right]\frac{\Gamma\left(1+l+s\right) \Gamma \left(1+l-s\right)}{\Gamma \left(1+2l\right)\Gamma \left(2l+2\right)}
\\
\times\left[1+(2 i M \omega-N)\psi(1+l+s)+(-2 i M \omega-N)\psi(1+l-s)+2N\psi(1+2l)+2N\psi(2+2l)\right]\,.
\end{multline}
Given the above final form for the dynamical tidal response function for generic spin perturbation of extremal Kerr BH, we can calculate the TLNs using the real part of it, which is (see \ref{app_B.2} for further details)
\begin{equation}\label{TLNs_ext_BH_s}
\,_{s}k_{lm}=(2m)^{2l-2N-2}\left(m^2\,_{s}\Tilde{k}^{(0)}_{lm}+ mM\omega
\,_{s}\Tilde{k}^{(1)}_{lm}\right) + \mathcal{O}(M^{2}\omega^{2})
\,.
\end{equation}
The explicit forms of $\,_{s}\Tilde{k}^{(0)}_{lm}$ and $\,_{s}\Tilde{k}^{(1)}_{lm}$ are given in \ref{app_B.2}. The above expression implies non-zero TLNs for generic perturbations associated with extremal Kerr BH. Interestingly, the TLNs are vanishing for axisymmetric tidal perturbation (see \ref{app_B.3} for other related details). This implies that both extremal as well as non-extremal BHs have vanishing TLNs under axisymmetric dynamical generic spin perturbations. Moreover, similar to the non-extremal case, for non-zero spin, the TLNs are not invariant under $s\to-s$ transformation. It remains to be seen if this has any connection to the electric and magnetic parts of the TLNs. 

\section{Conclusions}\label{section:6}

In this work, we studied the response of an arbitrarily rotating BH, which can be either non-extremal or extremal, under scalar, electromagnetic, and gravitational tidal fields. Our studies include both static and dynamical tidal fields. We have used the Teukolsky equation in the small frequency and near-zone regimes and calculated the response of a non-extremal Kerr BH to the external tidal field. Since the coordinate transformation defined for the calculation of the tidal response of a non-extremal Kerr BH was not well-defined for extremal Kerr BH, we analyzed the 
extremal case
separately. We also separately studied the response of a non-extremal Kerr BH to the scalar tidal field.

This work also generalizes the results found in Ref.~\cite{Bhatt:2024yyz}, where we studied the tidal response of an arbitrarily rotating BH in a gravitational tidal field with $s=-2$. The results found here can be summarized as follows:
\begin{itemize}
    \item Static tidal Love numbers of black holes vanish for generic scalar, electromagnetic, and gravitational tidal fields.
    \item Dynamical tidal Love numbers of black holes vanish for axisymmetric scalar, electromagnetic, and gravitational tidal fields.
    \item Dynamical tidal Love numbers of Schwarzschild and slowly rotating Kerr black holes also vanish for non-axisymmetric scalar, electromagnetic, and gravitational tidal fields.
    \item Dynamical tidal Love numbers of non-extremal and extremal Kerr black holes do not vanish for non-axisymmetric scalar, electromagnetic, and gravitational tidal fields, in general.
    \item The logarithmic contribution to the dynamical TLNs scales as $am\omega$ and exists for non-extremal Kerr black holes, with arbitrary spin, in non-axisymmetric electromagnetic and gravitational tidal fields.
\end{itemize}
We have tabulated these results in \ref{table_2}. Moreover, for non-axisymmetric electromagnetic and gravitational tidal fields, tidal Love numbers do not remain the same when one changes the sign of the spin-weight, i.e., under the transformation $s\to -s$. It is yet to be explored if it has any connection with the electric and magnetic tidal Love numbers. Thus, the non-zero dynamical TLNs found for extremal and non-extremal Kerr black holes in Ref.~\cite{Bhatt:2024yyz} are not only a characteristic of the gravitational tidal field but also of the scalar and the electromagnetic tidal fields.

\begin{table}[ht]
    \centering
    \setlength{\tabcolsep}{5pt}
    \renewcommand{\arraystretch}{1.0}
    \begin{tabular}{cccccc}
    \hline
        & Type of & $s$ & Schwarzschild & Non-extremal & Extremal\\
        & perturbation &  & and & Kerr BH & Kerr BH \\
        &  &  & slowly rotating &  &  \\
        &  &  & Kerr BHs &  &  \\\hline
         
       Static &Arbitrary &  Any & $0$ & $0$ & $0$\\
       $(\omega = 0)$ &  &   &  &  & \\ \hline
        
       Dynamic &Arbitrary &  Any & $0$ & $0$ & $0$\\
       $(\omega \ne 0,~m=0)$ & &  &  &  & \\ \hline
        
        &Scalar &  $~~0$ & $0$ & \ref{TLNs_non_extremal_Kerr_scalar_non_axi_tidal_field}; \ref{eq:scalar} & \ref{TLNs_ext_BH_s}; \ref{eq:Extremal_scalar_s_0}\\
        &Electromagnetic &  $~~1$ & $0$ & \ref{dynamic_TLNs}; \ref{eq:EMs+1} & \ref{TLNs_ext_BH_s}; \ref{eq:Extremal_EM_s_+1}\\ 
       Dynamic &Electromagnetic &  $-1$ & $0$ & \ref{dynamic_TLNs}; \ref{eq:EMs-1} & \ref{TLNs_ext_BH_s}; \ref{eq:Extremal_EM_s_-1}\\
       $(\omega \ne 0,~m\ne0)$ &Gravitational &  $~~2$ & $0$ & \ref{dynamic_TLNs}; \ref{eq:Gs+2} & \ref{TLNs_ext_BH_s}; \ref{eq:Extremal_G_s_+2}\\
        &Gravitational &  $-2$ & $0$ & \ref{dynamic_TLNs}; \ref{eq:Gs-2} & \ref{TLNs_ext_BH_s}; \ref{eq:Extremal_G_s_-2}\\ \hline
    \end{tabular}
\caption{Static and dynamical TLNs (without the logarithmic term) have been summarized for different types of BHs and their tidal perturbations. The logarithmic contributions to the dynamical TLNs are non-zero, only for $s=\pm 2$ and $s=-1$, and for Kerr BH (with arbitrary angular momentum, but within the extremal limit). For completeness, we note that, by slowly rotating Kerr BHs, we have implied that we consider only linear order terms in the angular momentum of the BH, neglecting second- and higher-order terms of the ratio $(a/M)$.}
\label{table_2}
\end{table}

This calculation can be extended to find  higher-order corrections in $M\omega$ to the expressions of TLNs derived here.
One can also pursue a scattering amplitude calculation for the gravitational and electromagnetic perturbations, just like the scalar ones studied for Schwarzschild BHs in Ref.~\cite{Creci:2021rkz}.
It will also be interesting to compare our results with those
obtained by using EFT and check if and where they match. Finally, seeking to confirm these tidal effects in real data would be another important test of the black hole solutions of General Relativity.

\section*{Acknowledgements}

Support from the National Science Foundation under Grant PHY-2309352 is acknowledged. Research of SC is supported by MATRICS (MTR/2023/000049) and Core Research (CRG/2023/000934) Grants from SERB, ANRF, Government of India. SC also acknowledges the warm hospitality at the Albert-Einstein Institute, Potsdam, where a part of this work was done. The visit was supported by a Max-Planck-India mobility grant. This document has been assigned the LIGO Preprint number LIGO-P2400610.

\appendix
\labelformat{section}{Appendix #1} 
\labelformat{subsection}{Appendix #1}
\section{Tidal response of a non-extremal Kerr black hole}\label{appendix_A}

In this Appendix, we shall explain the steps involved in simplifying the tidal response function of a non-extremal Kerr BH, presented in \ref{resp_func_arb_rot}. We will also describe the steps involved in the simplification of the expression of TLNs of a non-extremal Kerr BH. All these expressions in this Appendix correspond to the non-logarithmic term of the tidal response function and tidal Love numbers. Unlike the practice adopted in the main text, here we will drop $\textrm{``(non-log)''}$ from the superscript, for brevity.

\subsection{Simplification of the tidal response function}\label{app_A.1}

In this Appendix, we will describe the steps involved in getting \ref{response_1} from \ref{resp_func_arb_rot}. Recall that \ref{resp_func_arb_rot} showed that
\begin{equation}\label{resp_func_arb_rot_a}
    \,_{s}F_{lm}^{\rm Kerr} = \frac{ \Gamma \left(-2 l-N_1+N_2-1\right)\Gamma \left(l+2 i P_++N_1+1\right) \Gamma \left(l-s-N_2+1\right)}{\Gamma \left(-l-s-N_1\right) \Gamma \left(-l+2 i P_++N_2\right)\Gamma \left(2 l+N_1-N_2+1\right) }.
\end{equation}
Now, we can use the reflection formula for the gamma functions, $\Gamma(z)\Gamma(1-z)=\{\pi/\sin(\pi z)\}$, to write
\begin{equation}
    \frac{\Gamma \left(-2 l-N_1+N_2-1\right)}{\Gamma \left(-l-s-N_1\right)} = \alpha \frac{\Gamma \left(1+l+s+N_1\right)}{\Gamma \left(2 l+N_1-N_2+2\right)},
\end{equation}
and
\begin{equation}
    \frac{1}{\Gamma \left(-l+2 i P_++N_2\right)} = \xi\,\Gamma \left(1+l-2 i P_+-N_2\right)\,,
\end{equation}
where we have defined $\alpha$ in \ref{def_alpha_kappa} of the main text, and 
\begin{equation}\label{def_alpha_s_xi_s}
\xi = \frac{\sin(-l+2 i P_++N_2)\pi}{\pi}\,.
\end{equation}
Now, we can rewrite the tidal response function as
\begin{equation}\label{resp_func_arb_rot_1_s}
    \,_{s}F_{lm}^{\rm Kerr} = \alpha\xi\frac{ \Gamma \left(1+l+s+N_1\right)\Gamma \left(l+2 i P_++N_1+1\right) \Gamma \left(l-s-N_2+1\right)\Gamma \left(1+l-2 i P_+-N_2\right)}{\Gamma \left(2 l+N_1-N_2+2\right) \Gamma \left(2 l+N_1-N_2+1\right) }.
\end{equation}
Since we are working in the small frequency ($M\omega\ll 1)$ approximation, we can expand the expand the above expression up to linear orders of $M\omega$. To do so, we can use~\cite{abramowitz_stegun_1972}
\begin{equation}
\Gamma(f(z))=\Gamma(f(z_0)) + (z-z_0)\psi(f(z_0))\Gamma(f(z_0))\frac{\mathrm{d}f(z)}{\mathrm{d}z}\bigg|_{z=z_0} + \mathcal{O}[(z-z_0)^2]\,.
\end{equation}
Here, $\psi(z)$ is the digamma function defined as $\psi(z) = \Gamma'(z)/\Gamma(z)$, with $\Gamma'(z)$ being the first-order derivative of $\Gamma(z)$ with respect to $z$. Now, we can reexpress the tidal response function as
\begin{multline}\label{resp_func_arb_rot_1_expansion}
    \,_{s}F_{lm}^{\rm Kerr} = \alpha\xi\frac{ \Gamma \left(1+l+s\right)\Gamma \left(l+2 i P_++1\right) \Gamma \left(l-s+1\right)\Gamma \left(1+l-2 i P_+\right)}{\Gamma \left(2 l+2\right) \Gamma \left(2 l+1\right) }\\\times\left[1+N_1\psi(1 +l + s)+N_1\psi(1 + l+2 i P_+)- N_2\psi(1 + l - s)-N_2\psi(1 + l-2 i P_+)\right.\\\left.- (N_1 - N_2)\psi(2 + 2 l)- (N_1 - N_2)\psi(1 + 2 l)\right].
\end{multline}
Since we are working in the linear order of $M\omega$, we have neglected the second and higher orders of $M\omega$ in the above expression. Now, with the help of the identity
\begin{equation}
\Gamma(l+1+2iP_+)\Gamma(l+1-2iP_+)=\frac{2i\pi P_+}{\sin(2i\pi P_+)} \prod_{j=1}^{l} (j^2+4P_+^2)\,,
\end{equation}
where $l$ is an integer, the response function reduces to
\begin{multline}
    \,_{s}F_{lm}^{\rm Kerr} = \alpha\xi\frac{ \Gamma \left(1+l+s\right)\Gamma \left(l-s+1\right)}{\Gamma \left(2 l+2\right) \Gamma \left(2 l+1\right) }\frac{2i\pi P_+}{\sin(2i\pi P_+)} \prod_{j=1}^{l} (j^2+4P_+^2)\\\times\left[1+N_1\psi(1 +l + s)+N_1\psi(1 + l+2 i P_+)- N_2\psi(1 + l - s)-N_2\psi(1 + l-2 i P_+)\right.\\\left.- (N_1 - N_2)\psi(2 + 2 l)- (N_1 - N_2)\psi(1 + 2 l)\right]\,,
\end{multline}
which is \ref{response_1} of the main text [using $\kappa$ from \ref{def_alpha_kappa}].

\subsection{Simplification of the expression of tidal Love numbers}\label{app_A.2}

In this Appendix, we shall describe the steps involved in simplifying the expression of the dynamical TLNs of a non-extremal Kerr BH [\ref{eq_39_a})]. We will start this simplification by separating the real and imaginary part of $N_1$ and $N_2$, given in \ref{def_N1} and \ref{def_N2}, respectively,
\begin{equation}\label{ReImN1}
\text{Re}(N_1) = am\omega\,x_{N_1}, \qquad \text{Im}(N_1) = M\omega\,y_{N_1}\,,
\end{equation}
and
\begin{equation}\label{ReImN2}
\text{Re}(N_2) = am\omega\,x_{N_2}, \qquad \text{Im}(N_2) = M\omega\,y_{N_2}\,,
\end{equation}
where
\begin{align}
x_{N_1}&=-\frac{2(l^2+l+s^2)}{l(l+1)(2l+1)}+\frac{4 (1-s) (r_+-M)^2}{a^2m^2+(r_+-M)^2s^2}\,,
\label{ReN1_a}
\\
y_{N_1}&=\frac{2}{M} \left[\frac{(4 l +2s) M-(2 l +2s-1)r_+}{2 l+1}-\frac{2 \left(r_+-M\right) \left\{-a^2m^2-(r_+-M)^2s\right\}}{a^2m^2+(r_+-M)^2s^2}\right]\,,
\label{ImN1_a}
\\
x_{N_2}&=\frac{2(l^2+l+s^2)}{l(l+1)(2l+1)}+\frac{4 (1-s) (r_+-M)^2}{a^2m^2+(r_+-M)^2s^2}\,,
\label{ReN2_a}
\\
y_{N_2}&=  \frac{2}{M} \left[\frac{(4l-2s+4)M - (2 l-2s+3) r_+}{2 l+1}-\frac{2 \left(r_+-M\right) \left\{-a^2m^2-(r_+-M)^2s\right\}}{a^2m^2+(r_+-M)^2s^2}\right]\,.
\label{ImN2_a}
\end{align}
Since we also have $(N_1-N_2)$ in the expression of the tidal Love numbers, we will also require the real and imaginary parts of $(N_1-N_2)$,
\begin{equation}\label{ReImN1mN2}
\text{Re}(N_1-N_2) = am\omega\,x_{N_{12}}\,, \qquad \text{Im}(N_1-N_2) = M\omega\,y_{N_{12}}\,,
\end{equation}
where
\begin{equation}\label{ReImN1mN2_a}
x_{N_{12}}=-\frac{4 (l^2+l+s^2)}{l(l+1)(2l+1)}\,, \qquad y_{N_{12}}=\frac{8 (1-s) \left(r_+-M\right)}{(2 l+1)M}\,.
\end{equation}
From the above expression, it is apparent that the real parts of $N_1$, $N_2$, and $(N_1-N_2)$ have $am\omega$ as a multiplication factor. In addition, the imaginary parts of $N_1$, $N_2$, and $(N_1-N_2)$ have $M\omega$ as the multiplication factor. 

Since the expression of the tidal Love numbers contain $\{N_1/(N_1-N_2)\}$, we calculate the real and imaginary parts of it:
\begin{equation}\label{Re_ratio}
\text{Re}\left(\frac{N_1}{N_1-N_2}\right)=\frac{(ma/M)^2\,x_{N_1}x_{N_{12}}+y_{N_1}y_{N_{12}}}{\{(ma/M)\,x_{N_{12}}\}^2+\{y_{N_{12}}\}^2}\,,
\end{equation}
and
\begin{equation}\label{Im_ratio}
\text{Im}\left(\frac{N_1}{N_1-N_2}\right) =\frac{ma}{M}\left[\frac{y_{N_1}x_{N_{12}}-x_{N_1}y_{N_{12}}}{\{(ma/M)\,x_{N_{12}}\}^2+\{y_{N_{12}}\}^2}\right]\,.
\end{equation}
Hence, the imaginary part of $\{N_1/(N_1-N_2)\}$ has $am$ as the multiplication factor. 

Now, we can simplify the expression of the TLNs [\ref{eq_39_a}] by using the above written expression and neglecting the second- and higher-order terms of $M\omega$. We will first follow this procedure in the expression inside the square brackets in \ref{eq_39_a}. It yields
\begin{align}
\,_{s}k_{lm}&=-(-1)^s P_{+}F_{2}\Bigg[-\pi\coth(2 P_+\pi)\text{Re}\left(\frac{N_1}{N_1-N_2}\right)\text{Re}N_{2}
+\text{Im}\left(\frac{N_1}{N_1-N_2}\right)\left\{1+\pi \coth(2 P_+\pi)\textrm{Im}N_{2}\right\}
\nonumber
\\
&+\text{Re}\left(\frac{N_1}{N_1-N_2}\right)\text{Im}(M\omega F_1)
+\text{Im}\left(\frac{N_1}{N_1-N_2}\right)\text{Re}(M\omega F_1)\Bigg]~.
\end{align}
Since $P_+$ also contains $M\omega$, we can ignore some more terms of second and higher orders of $M\omega$. After dropping those terms, we get
\begin{align}
\,_{s}k_{lm}&=-(-1)^s \left(\frac{am}{r_+-r_-}\right)F_{2}\text{Im}\left(\frac{N_1}{N_1-N_2}\right)\nonumber
\\
&-(-1)^s \left(\frac{am\omega}{r_+-r_-}\right)F_{2}\Bigg[\left\{-\pi\coth(2 P_+\pi)\frac{\text{Re}N_{2}}{\omega}+\frac{\text{Im}(M\omega F_1)}{\omega}\right\}\text{Re}\left(\frac{N_1}{N_1-N_2}\right)
\nonumber
\\
&+\left\{\pi \coth(2 P_+\pi)\frac{\textrm{Im}N_{2}}{\omega}+\frac{\text{Re}(M\omega F_1)}{\omega}-\frac{2Mr_+}{am}\right\}\text{Im}\left(\frac{N_1}{N_1-N_2}\right)\Bigg]~.
\end{align}
Now, we can use the above written expression of the real and imaginary parts of $N_1$, $N_2$, and $\{N_1/(N_1-N_2)\}$ and write the expression of the TLNs as
\begin{align}
\,_{s}k_{lm}&=-(-1)^s (am)^2\frac{F_{2}}{(r_+-r_-)M}\frac{y_{N_1}x_{N_{12}}-x_{N_1}y_{N_{12}}}{\{(ma/M)\,x_{N_{12}}\}^2+\{y_{N_{12}}\}^2}
\nonumber
\\
&-(-1)^s am\omega\frac{F_2}{r_+-r_-}\Bigg[\left\{-am\,x_{N_2}\pi\coth(2 P_+\pi)+\frac{\text{Im}(M\omega F_1)}{\omega}\right\}\frac{(ma/M)^2\,x_{N_1}x_{N_{12}}+y_{N_1}y_{N_{12}}}{\{(ma/M)\,x_{N_{12}}\}^2+\{y_{N_{12}}\}^2}
\nonumber
\\
&+\left\{am\,y_{N_2}\pi \coth(2 P_+\pi)+\frac{ma}{M}\frac{\text{Re}(M\omega F_1)}{\omega}-2r_+\right\}\,\frac{y_{N_1}x_{N_{12}}-x_{N_1}y_{N_{12}}}{\{(ma/M)\,x_{N_{12}}\}^2+\{y_{N_{12}}\}^2}\Bigg]~.
\end{align}
We can further simplify the above expression by ignoring the second- and higher-order terms arising from the contribution due to $F_2$, $P_+$, and $M\omega F_1$. This yields
\begin{align}
\,_{s}k_{lm}&=-(-1)^s (am)^2\frac{\Tilde{F}_{2}}{(r_+-r_-)M}\frac{y_{N_1}x_{N_{12}}-x_{N_1}y_{N_{12}}}{\{(ma/M)\,x_{N_{12}}\}^2+\{y_{N_{12}}\}^2}
\nonumber
\\
&-(-1)^s am\omega\frac{\Tilde{F}_2}{r_+-r_-}\Bigg[\left\{-am\,x_{N_2}\pi\coth(2 P_+^A\pi)+\frac{\text{Im}(M\omega \Tilde{F}_1)}{\omega}\right\}\frac{(ma/M)^2\,x_{N_1}x_{N_{12}}+y_{N_1}y_{N_{12}}}{\{(ma/M)\,x_{N_{12}}\}^2+\{y_{N_{12}}\}^2}
\nonumber
\\
&+\Bigg\{am\,y_{N_2}\pi \coth(2 P_+^A\pi)+\frac{ma}{M}\frac{\text{Re}(M\omega \Tilde{F}_1)}{\omega}-2r_+\nonumber
\\
&-\frac{4amr_+}{(r_+-r_-)}\sum_{k=1}^{l}\frac{4P_+^A}{k^2+4(P_+^A)^2}\Bigg\}\,\frac{y_{N_1}x_{N_{12}}-x_{N_1}y_{N_{12}}}{\{(ma/M)\,x_{N_{12}}\}^2+\{y_{N_{12}}\}^2}\Bigg]~,
\end{align}
where
\begin{equation}
P_+^A = \frac{am}{r_+-r_-}, \qquad \Tilde{F}_2 = \frac{ \Gamma \left(1+l+s\right)\Gamma \left(l-s+1\right)}{\Gamma \left(2 l+2\right) \Gamma \left(2 l+1\right) }\prod_{j=1}^{l} \left[j^2+4(P_+^A)^2\right]\,,
\end{equation}
and 
\begin{multline}
M\omega \Tilde{F}_1=N_1\psi(1 +l + s)+N_1\psi(1 + l+2 i P_+^A)- N_2\psi(1 + l - s)-N_2\psi(1 + l-2 i P_+^A)\\
- (N_1 - N_2)\psi(2 + 2 l)- (N_1 - N_2)\psi(1 + 2 l)\,.
\end{multline}
Hence, we can rewrite the dynamical TLNs as \ref{dynamic_TLNs} of the main text, with $\,_{s}k^{(0)}_{lm}$ and $\,_{s}k^{(1)}_{lm}$ being given by
\begin{equation}
\,_{s}k^{(0)}_{lm}=-(-1)^s \frac{\Tilde{F}_{2}}{(r_+-r_-)M}\frac{y_{N_1}x_{N_{12}}-x_{N_1}y_{N_{12}}}{\{(ma/M)\,x_{N_{12}}\}^2+\{y_{N_{12}}\}^2}\,,
\end{equation}
and
\begin{align}
\,_{s}k^{(1)}_{lm}&=-(-1)^s \frac{\Tilde{F}_2}{r_+-r_-}\Bigg[\left\{-am\,x_{N_2}\pi\coth(2 P_+^A\pi)+\frac{\text{Im}(M\omega \Tilde{F}_1)}{\omega}\right\}\frac{(ma/M)^2\,x_{N_1}x_{N_{12}}+y_{N_1}y_{N_{12}}}{\{(ma/M)\,x_{N_{12}}\}^2+\{y_{N_{12}}\}^2}
\nonumber
\\
&+\Bigg\{am\,y_{N_2}\pi \coth(2 P_+^A\pi)+\frac{ma}{M}\frac{\text{Re}(M\omega \Tilde{F}_1)}{\omega}-2r_+
\\
\nonumber
&-\frac{4amr_+}{(r_+-r_-)}\sum_{k=1}^{l}\frac{4P_+^A}{k^2+4(P_+^A)^2}\Bigg\}\,\frac{y_{N_1}x_{N_{12}}-x_{N_1}y_{N_{12}}}{\{(ma/M)\,x_{N_{12}}\}^2+\{y_{N_{12}}\}^2}\Bigg]~.
\end{align}
\subsection{Tidal Love numbers of a non-extremal Kerr black hole for different tidal perturbations}
In this Appendix, we will list the values $\,_{s}k^{(0)}_{lm}$ and $\,_{s}k^{(1)}_{lm}$ of TLNs for electromagnetic and gravitational tidal perturbations:
\begin{itemize}
    \item {\bf Electromagnetic perturbation} $\bm{(s=+1)}$
\begin{subequations}\label{eq:EMs+1}
\begin{equation}
\,_{1}k^{(0)}_{lm}= \frac{\Tilde{F}_{2}}{(r_+-r_-)M}\frac{y_{N_1}x_{N_{12}}-x_{N_1}y_{N_{12}}}{\{(ma/M)\,x_{N_{12}}\}^2+\{y_{N_{12}}\}^2}\,,
\end{equation}
and
\begin{align}
\,_{1}k^{(1)}_{lm}&= \frac{\Tilde{F}_2}{r_+-r_-}\Bigg[\left\{-am\,x_{N_2}\pi\coth(2 P_+^A\pi)+\frac{\text{Im}(M\omega \Tilde{F}_1)}{\omega}\right\}\frac{(ma/M)^2\,x_{N_1}x_{N_{12}}+y_{N_1}y_{N_{12}}}{\{(ma/M)\,x_{N_{12}}\}^2+\{y_{N_{12}}\}^2}
\nonumber
\\
&+\Bigg\{am\,y_{N_2}\pi \coth(2 P_+^A\pi)+\frac{ma}{M}\frac{\text{Re}(M\omega \Tilde{F}_1)}{\omega}-2r_+
\\
\nonumber
&-\frac{4amr_+}{(r_+-r_-)}\sum_{k=1}^{l}\frac{4P_+^A}{k^2+4(P_+^A)^2}\Bigg\}\,\frac{y_{N_1}x_{N_{12}}-x_{N_1}y_{N_{12}}}{\{(ma/M)\,x_{N_{12}}\}^2+\{y_{N_{12}}\}^2}\Bigg]~,
\end{align}
where
\begin{equation}
P_+^A = \frac{am}{r_+-r_-}, \qquad \Tilde{F}_2 = \frac{ \Gamma \left(2+l\right)\Gamma \left(l\right)}{\Gamma \left(2 l+2\right) \Gamma \left(2 l+1\right) }\prod_{j=1}^{l} \left[j^2+4(P_+^A)^2\right]\,,
\end{equation}
\begin{multline}
M\omega \Tilde{F}_1=N_1\psi(2 +l)+N_1\psi(1 + l+2 i P_+^A)- N_2\psi(l)-N_2\psi(1 + l-2 i P_+^A)\\
- (N_1 - N_2)\psi(2 + 2 l)- (N_1 - N_2)\psi(1 + 2 l)\,,
\end{multline}
with
\begin{equation}
N_{1}=2\omega\left[-\frac{a m(l^2+l+1)}{l(l+1)(2l+1)}+ir_+\right]~, \qquad N_{2}=2\omega\left[\frac{a m(l^2+l+1)}{l(l+1)(2l+1)}+ir_+\right]~,
\end{equation}
\begin{equation}
x_{N_1}=-\frac{2(l^2+l+1)}{l(l+1)(2l+1)}, \qquad y_{N_1}=\frac{2r_+}{M}\,,
\end{equation}

\begin{equation}
    x_{N_2}=\frac{2(l^2+l+1)}{l(l+1)(2l+1)}\,, \qquad y_{N_2}=  \frac{2r_+}{M}\,,
\end{equation}

and
\begin{equation}
x_{N_{12}}=-\frac{4 (l^2+l+1)}{l(l+1)(2l+1)}\,, \qquad y_{N_{12}}=0\,.
\end{equation}
\end{subequations}

\item {\bf Electromagnetic perturbation} $\bm{(s=-1)}$
\begin{subequations}\label{eq:EMs-1}
\begin{equation}
\,_{-1}k^{(0)}_{lm}=\frac{\Tilde{F}_{2}}{(r_+-r_-)M}\frac{y_{N_1}x_{N_{12}}-x_{N_1}y_{N_{12}}}{\{(ma/M)\,x_{N_{12}}\}^2+\{y_{N_{12}}\}^2}\,,
\end{equation}
and
\begin{align}
\,_{-1}k^{(1)}_{lm}&=\frac{\Tilde{F}_2}{r_+-r_-}\Bigg[\left\{-am\,x_{N_2}\pi\coth(2 P_+^A\pi)+\frac{\text{Im}(M\omega \Tilde{F}_1)}{\omega}\right\}\frac{(ma/M)^2\,x_{N_1}x_{N_{12}}+y_{N_1}y_{N_{12}}}{\{(ma/M)\,x_{N_{12}}\}^2+\{y_{N_{12}}\}^2}
\nonumber
\\
&+\Bigg\{am\,y_{N_2}\pi \coth(2 P_+^A\pi)+\frac{ma}{M}\frac{\text{Re}(M\omega \Tilde{F}_1)}{\omega}-2r_+
\\
\nonumber
&-\frac{4amr_+}{(r_+-r_-)}\sum_{k=1}^{l}\frac{4P_+^A}{k^2+4(P_+^A)^2}\Bigg\}\,\frac{y_{N_1}x_{N_{12}}-x_{N_1}y_{N_{12}}}{\{(ma/M)\,x_{N_{12}}\}^2+\{y_{N_{12}}\}^2}\Bigg]~,
\end{align}
where
\begin{equation}
P_+^A = \frac{am}{r_+-r_-}, \qquad \Tilde{F}_2 = \frac{ \Gamma \left(l\right)\Gamma \left(2+l\right)}{\Gamma \left(2 l+2\right) \Gamma \left(2 l+1\right) }\prod_{j=1}^{l} \left[j^2+4(P_+^A)^2\right]\,,
\end{equation}
\begin{multline}
M\omega \Tilde{F}_1=N_1\psi(l)+N_1\psi(1 + l+2 i P_+^A)- N_2\psi(2 + l)-N_2\psi(1 + l-2 i P_+^A)\\
- (N_1 - N_2)\psi(2 + 2 l)- (N_1 - N_2)\psi(1 + 2 l)\,,
\end{multline}
with

\begin{multline}
N_{1}=2\omega\left[-\frac{a m}{l(l+1)(2l+1)}+\frac{i \left(i a m+4 l M-2 l r_++r_+\right)}{2 l+1}\right.\\\left.+\frac{2 \left(M-r_+\right) \left(-ia m+M-r_+\right)}{\left(a m-i M +i r_+\right)}-\frac{2 i \left(M-r_+\right)}{2 l+1}\right]~,
\end{multline}
\begin{multline}
N_{2}=2\omega\left[\frac{a m}{l(l+1)(2l+1)}+\frac{a m+4 i (l+1) M-i (2 l+3) r_+}{2 l+1}\right.\\\left.+\frac{2 \left(M-r_+\right) \left(-i a m+M-r_+\right)}{\left(a m-i M+i r_+\right)}+\frac{2 i \left(M-r_+\right)}{2 l+1}\right]~,
\end{multline}
\begin{equation}
x_{N_1}=-\frac{2(l^2+l+1)}{l(l+1)(2l+1)}+\frac{8 (r_+-M)^2}{a^2m^2+(r_+-M)^2}\,,
\end{equation}

\begin{equation}
y_{N_1}=\frac{2}{M} \left[\frac{(4 l -2) M-(2 l -3)r_+}{2 l+1}-\frac{2 \left(r_+-M\right) \left\{-a^2m^2+(r_+-M)^2\right\}}{a^2m^2+(r_+-M)^2}\right]\,,
\end{equation}

\begin{equation}
    x_{N_2}=\frac{2(l^2+l+1)}{l(l+1)(2l+1)}+\frac{8 (r_+-M)^2}{a^2m^2+(r_+-M)^2}\,,
\end{equation}

\begin{equation}
y_{N_2}=  \frac{2}{M} \left[\frac{2(2l+3)M - (2 l+5) r_+}{2 l+1}-\frac{2 \left(r_+-M\right) \left\{-a^2m^2+(r_+-M)^2\right\}}{a^2m^2+(r_+-M)^2}\right]\,,
\end{equation}
and
\begin{equation}
x_{N_{12}}=-\frac{4 (l^2+l+1)}{l(l+1)(2l+1)}\,, \qquad y_{N_{12}}=\frac{16 \left(r_+-M\right)}{(2 l+1)M}\,.
\end{equation}
\end{subequations}

\item {\bf Gravitational perturbation} $\bm{(s=+2)}$
\begin{subequations}\label{eq:Gs+2}
\begin{equation}
\,_{2}k^{(0)}_{lm}=- \frac{\Tilde{F}_{2}}{(r_+-r_-)M}\frac{y_{N_1}x_{N_{12}}-x_{N_1}y_{N_{12}}}{\{(ma/M)\,x_{N_{12}}\}^2+\{y_{N_{12}}\}^2}\,,
\end{equation}
and
\begin{align}
\,_{2}k^{(1)}_{lm}&=- \frac{\Tilde{F}_2}{r_+-r_-}\Bigg[\left\{-am\,x_{N_2}\pi\coth(2 P_+^A\pi)+\frac{\text{Im}(M\omega \Tilde{F}_1)}{\omega}\right\}\frac{(ma/M)^2\,x_{N_1}x_{N_{12}}+y_{N_1}y_{N_{12}}}{\{(ma/M)\,x_{N_{12}}\}^2+\{y_{N_{12}}\}^2}
\nonumber
\\
&+\Bigg\{am\,y_{N_2}\pi \coth(2 P_+^A\pi)+\frac{ma}{M}\frac{\text{Re}(M\omega \Tilde{F}_1)}{\omega}-2r_+
\\
\nonumber
&-\frac{4amr_+}{(r_+-r_-)}\sum_{k=1}^{l}\frac{4P_+^A}{k^2+4(P_+^A)^2}\Bigg\}\,\frac{y_{N_1}x_{N_{12}}-x_{N_1}y_{N_{12}}}{\{(ma/M)\,x_{N_{12}}\}^2+\{y_{N_{12}}\}^2}\Bigg]~,
\end{align}
where
\begin{equation}
P_+^A = \frac{am}{r_+-r_-}, \qquad \Tilde{F}_2 = \frac{ \Gamma \left(3+l\right)\Gamma \left(l-1\right)}{\Gamma \left(2 l+2\right) \Gamma \left(2 l+1\right) }\prod_{j=1}^{l} \left[j^2+4(P_+^A)^2\right]\,,
\end{equation}
\begin{multline}
M\omega \Tilde{F}_1=N_1\psi(3 +l)+N_1\psi(1 + l+2 i P_+^A)- N_2\psi(l - 1)-N_2\psi(1 + l-2 i P_+^A)\\
- (N_1 - N_2)\psi(2 + 2 l)- (N_1 - N_2)\psi(1 + 2 l)\,,
\end{multline}
with

\begin{multline}
N_{1}=2\omega\left[-\frac{4 a m}{l(l+1)(2l+1)}+\frac{i \left(i a m+4 l M-2 l r_++r_+\right)}{2 l+1}\right.\\\left.+\frac{2 \left(M-r_+\right) \left(-ia m+M-r_+\right)}{\left(a m+2i M-2i r_+\right)}+\frac{4 i \left(M-r_+\right)}{2 l+1}\right]~,
\end{multline}
\begin{multline}
N_{2}=2\omega\left[\frac{4 a m}{l(l+1)(2l+1)}+\frac{a m+4 i (l+1) M-i (2 l+3) r_+}{2 l+1}\right.\\\left.+\frac{2 \left(M-r_+\right) \left(-i a m+M-r_+\right)}{\left(a m+2i M -2i r_+\right)}-\frac{4 i \left(M-r_+\right)}{2 l+1}\right]~,
\end{multline}
\begin{equation}
x_{N_1}=-\frac{2(l^2+l+4)}{l(l+1)(2l+1)}-\frac{4 (r_+-M)^2}{a^2m^2+4(r_+-M)^2}\,,
\end{equation}

\begin{equation}
y_{N_1}=\frac{2}{M} \left[\frac{4(l +1) M-(2 l +3)r_+}{2 l+1}-\frac{2 \left(r_+-M\right) \left\{-a^2m^2-2(r_+-M)^2\right\}}{a^2m^2+4(r_+-M)^2}\right]\,,
\end{equation}

\begin{equation}
    x_{N_2}=\frac{2(l^2+l+4)}{l(l+1)(2l+1)}-\frac{4 (r_+-M)^2}{a^2m^2+4(r_+-M)^2}\,,
\end{equation}

\begin{equation}
y_{N_2}=  \frac{2}{M} \left[\frac{4lM - (2 l-1) r_+}{2 l+1}-\frac{2 \left(r_+-M\right) \left\{-a^2m^2-2(r_+-M)^2\right\}}{a^2m^2+4(r_+-M)^2}\right]\,,
\end{equation}
and
\begin{equation}
x_{N_{12}}=-\frac{4 (l^2+l+4)}{l(l+1)(2l+1)}\,, \qquad y_{N_{12}}=-\frac{8 \left(r_+-M\right)}{(2 l+1)M}\,.
\end{equation}
\end{subequations}

\item {\bf Gravitational perturbation} $\bm{(s=-2)}$
\begin{subequations}\label{eq:Gs-2}
\begin{equation}
\,_{-2}k^{(0)}_{lm}=- \frac{\Tilde{F}_{2}}{(r_+-r_-)M}\frac{y_{N_1}x_{N_{12}}-x_{N_1}y_{N_{12}}}{\{(ma/M)\,x_{N_{12}}\}^2+\{y_{N_{12}}\}^2}\,,
\end{equation}
and
\begin{align}
\,_{-2}k^{(1)}_{lm}&=- \frac{\Tilde{F}_2}{r_+-r_-}\Bigg[\left\{-am\,x_{N_2}\pi\coth(2 P_+^A\pi)+\frac{\text{Im}(M\omega \Tilde{F}_1)}{\omega}\right\}\frac{(ma/M)^2\,x_{N_1}x_{N_{12}}+y_{N_1}y_{N_{12}}}{\{(ma/M)\,x_{N_{12}}\}^2+\{y_{N_{12}}\}^2}
\nonumber
\\
&+\Bigg\{am\,y_{N_2}\pi \coth(2 P_+^A\pi)+\frac{ma}{M}\frac{\text{Re}(M\omega \Tilde{F}_1)}{\omega}-2r_+
\\
\nonumber
&-\frac{4amr_+}{(r_+-r_-)}\sum_{k=1}^{l}\frac{4P_+^A}{k^2+4(P_+^A)^2}\Bigg\}\,\frac{y_{N_1}x_{N_{12}}-x_{N_1}y_{N_{12}}}{\{(ma/M)\,x_{N_{12}}\}^2+\{y_{N_{12}}\}^2}\Bigg]~,
\end{align}
where
\begin{equation}
P_+^A = \frac{am}{r_+-r_-}, \qquad \Tilde{F}_2 = \frac{ \Gamma \left(l-1\right)\Gamma \left(3+l\right)}{\Gamma \left(2 l+2\right) \Gamma \left(2 l+1\right) }\prod_{j=1}^{l} \left[j^2+4(P_+^A)^2\right]\,,
\end{equation}
\begin{multline}
M\omega \Tilde{F}_1=N_1\psi(l -1)+N_1\psi(1 + l+2 i P_+^A)- N_2\psi(3 + l)-N_2\psi(1 + l-2 i P_+^A)\\
- (N_1 - N_2)\psi(2 + 2 l)- (N_1 - N_2)\psi(1 + 2 l)\,,
\end{multline}
with

\begin{multline}
N_{1}=2\omega\left[-\frac{4 a m}{l(l+1)(2l+1)}+\frac{i \left(i a m+4 l M-2 l r_++r_+\right)}{2 l+1}\right.\\\left.+\frac{2 \left(M-r_+\right) \left(-ia m+M-r_+\right)}{\left(a m-2i M+2i r_+\right)}-\frac{4 i \left(M-r_+\right)}{2 l+1}\right]~,
\end{multline}
\begin{multline}
N_{2}=2\omega\left[\frac{4 a m}{l(l+1)(2l+1)}+\frac{a m+4 i (l+1) M-i (2 l+3) r_+}{2 l+1}\right.\\\left.+\frac{2 \left(M-r_+\right) \left(-i a m+M-r_+\right)}{\left(a m-2i M+2i r_+\right)}+\frac{4 i \left(M-r_+\right)}{2 l+1}\right]~,
\end{multline}
\begin{equation}
x_{N_1}=-\frac{2(l^2+l+4)}{l(l+1)(2l+1)}+\frac{12 (r_+-M)^2}{a^2m^2+4(r_+-M)^2}\,,
\end{equation}

\begin{equation}
y_{N_1}=\frac{2}{M} \left[\frac{4(l-1) M-(2 l -5)r_+}{2 l+1}-\frac{2 \left(r_+-M\right) \left\{-a^2m^2+2(r_+-M)^2\right\}}{a^2m^2+4(r_+-M)^2}\right]\,,
\end{equation}

\begin{equation}
    x_{N_2}=\frac{2(l^2+l+4)}{l(l+1)(2l+1)}+\frac{12 (r_+-M)^2}{a^2m^2+4(r_+-M)^2}\,,
\end{equation}

\begin{equation}
y_{N_2}=  \frac{2}{M} \left[\frac{4(l+2)M - (2 l+7) r_+}{2 l+1}-\frac{2 \left(r_+-M\right) \left\{-a^2m^2+2(r_+-M)^2\right\}}{a^2m^2+4(r_+-M)^2}\right]\,,
\end{equation}
and
\begin{equation}
x_{N_{12}}=-\frac{4 (l^2+l+4)}{l(l+1)(2l+1)}\,, \qquad y_{N_{12}}=\frac{24 \left(r_+-M\right)}{(2 l+1)M}\,.
\end{equation}
\end{subequations}
\end{itemize}
\section{Tidal Love numbers of a non-extremal Kerr black hole to the non-axisymmetric scalar tidal field}\label{appendix_B_New}
In this Appendix, we will provide the expressions of $\,_{0}k_{lm}^{(0)}$ and $\,_{0}k_{lm}^{(1)}$ from \ref{TLNs_non_extremal_Kerr_scalar_non_axi_tidal_field}. Similar to the last Appendix, we can simplify \ref{TLNs_non_extremal_Kerr_scalar_non_axi_tidal_field_0} and arrive at \ref{TLNs_non_extremal_Kerr_scalar_non_axi_tidal_field}, where
\begin{subequations}\label{eq:scalar}
\begin{equation}
    \,_{0}k_{lm}^{(0)} = \frac{(2l+1)r_+}{2(r_+-r_-)}\Tilde{F}_2,
\end{equation}
and
\begin{align}
    \,_{0}k_{lm}^{(1)} &= \frac{(2l+1)r_+}{(r_+-r_-)}\frac{P_+^B}{M\omega}\Tilde{F}_2\sum_{k=1}^{l}\frac{4P_+^A}{k^2+4(P_+^A)^2}
-\frac{1}{2}\left(\frac{(a/M)m}{r_+-r_-}\right)\Tilde{F}_2\left\{-\pi\coth(2 P_+\pi)am\,x_{U_2}+\frac{\text{Im}(M\omega\Tilde{F}_1)}{\omega}\right\}
\nonumber
\\
&+\frac{(2l+1)r_+}{2M(r_+-r_-)}\Tilde{F}_2\left\{\pi \coth(2 P_+\pi)M\,y_{U_2}+\frac{\text{Re}(M\omega\Tilde{F}_1)}{\omega}\right\}-\frac{(2l+1)r_+^2}{M(r_+-r_-)}\Tilde{F}_2\frac{M}{am},
\end{align}
with
\begin{equation}
    \Tilde{F}_2 = \frac{ \Gamma \left(1+l\right)\Gamma \left(1+l\right)}{\Gamma \left(2 l+2\right) \Gamma \left(2 l+1\right) }\prod_{j=1}^{l} \left[j^2+4(P_+^A)^2\right], \qquad P_+^A = \frac{am}{r_+-r_-}, \qquad P_+^B = -\frac{2Mr_+\omega}{r_+-r_-},
\end{equation}
and
\begin{align}\label{def_Mo_F1_2}
M\omega\Tilde{F}_1\equiv U_1\psi(1 +l)&+U_1\psi(1 + l+2 i P_+^A)
- U_2\psi(1 + l)-U_2\psi(1 + l-2 i P_+^A)
\nonumber 
\\
&- (U_1 - U_2)\psi(2 + 2 l)- (U_1 - U_2)\psi(1 + 2 l)\,.
\end{align}
In addition,
\begin{equation}
U_1 = 2ir_+\omega - \frac{2am\omega}{2l+1}\,,
\qquad 
U_2 = 2ir_+\omega + \frac{2am\omega}{2l+1}\,, \qquad x_{U_2}=\frac{2}{2l+1}\,, \qquad y_{U_2}=  \frac{2r_+}{M}\,
\end{equation}
\end{subequations}
\section{Tidal response of an extremal Kerr black hole}\label{appendix_B}

In this Appendix, we will chalk out the steps involved in the simplification the tidal response function of an extremal Kerr BH [\ref{non_static_rf_ext_BH}]. We will also explain the steps involved in arriving the expression of the TLNs from the tidal response function [\ref{simplified_non_static_rf_ext_BH_02}]. In addition, we will discuss the specific case of the axisymmetric tidal perturbation for an extremal Kerr BH.

\subsection{Simplification of the tidal response function for an extremal Kerr black hole}\label{app_B.1}

In this Appendix, we will explain the steps involved in arriving at \ref{simplified_non_static_rf_ext_BH} from \ref{non_static_rf_ext_BH}. The tidal response function is
\begin{align}
\,_{s}F_{lm}&=-\{2i(m-2M\omega)\}^{2l+1-2N}\frac{\Gamma \left(-2l+2N\right) \Gamma \left(1+l-s-2 i M \omega-N\right)}{\Gamma \left(-l-s-2 i M \omega+N\right)\Gamma \left(2l+2 -2N\right)}\,.
\end{align}
Using the reflection formula for the Gamma functions, we can write
\begin{equation}
\frac{\Gamma \left(-2l+2N\right)}{\Gamma \left(-l-s-2 i M \omega+N\right)} = \frac{\sin(l+s+2 i M \omega-N)\pi}{\sin(2l-2N)\pi}\frac{\Gamma \left(1+l+s+2 i M \omega-N\right)}{\Gamma \left(1+2l-2N\right)}\,.
\end{equation}
For $l\in \mathbb{Z}_{\ge0}$, $s\in\mathbb{Z}$, $l\ge |s|$, and $M\omega \neq 0$,
\begin{equation}
\frac{\sin(l+s+2 i M \omega-N)\pi}{\sin(2l-2N)\pi}=-(-1)^{l+s}\frac{\sin(2 i M \omega-N)\pi}{\sin(2N)\pi}\,.
\end{equation}
It implies
\begin{multline}
\,_{s}F_{lm} = (-1)^{l+s}i^{2l+1-2N}\{2(m-2M\omega)\}^{2l+1-2N}\frac{\sin(2 i M \omega-N)\pi}{\sin(2N)\pi}
\\
\times\frac{\Gamma \left(1+l+s+2 i M \omega-N\right) \Gamma \left(1+l-s-2 i M \omega-N\right)}{\Gamma \left(1+2l-2N\right)\Gamma \left(2l+2 -2N\right)}\,.
\end{multline}
As done in the last Appendix, we can expand the above expression in the linear order of $M\omega$ and neglect the second- and higher-order terms of $M\omega$. This leads to \ref{simplified_non_static_rf_ext_BH} of the main text.

\subsection{Tidal Love numbers of an extremal Kerr black hole}\label{app_B.2}
In this section, we will chalk out the steps involved in arriving at the TLNs [\ref{TLNs_ext_BH_s}] from the tidal response function [\ref{simplified_non_static_rf_ext_BH_02}]. Since TLNs are related to the real part of the tidal response function, we can write
\begin{align}\label{tlnextremalbh}
\,_{s}k_{lm}&=\frac{1}{2}\textrm{Re} \,_{s}F_{lm}=\frac{(-1)^s}{2} (2m)^{2l-2N-1}\left\{m-(2l+1)2M\omega\right\}\left(\frac{\Gamma\left(1+l+s\right) \Gamma \left(1+l-s\right)}{\Gamma \left(1+2l\right)\Gamma \left(2l+2\right)}\right)
\nonumber
\\
&\times \Bigg[-\frac{2l(l+1)(2l+1)}{\left(l^2+l+s^2\right)}\left\{1+N\left(2\psi(1+2l)+2\psi(2+2l)-\psi(1+l+s)-\psi(1+l-s)\right) \right\}
\nonumber
\\
&\qquad \qquad +4Mm\omega \left\{\psi(1+l+s)-\psi(1+l-s)-\frac{\pi m \left(l^2+l+s^2\right)}{l(l+1)(2l+1)}\right\}\Bigg]\,.
\end{align}
Now, we can further simplify the above expression and write it as \ref{TLNs_ext_BH_s}, where
\begin{equation}
    \,_{s}\Tilde{k}^{(0)}_{lm} = -\frac{(-1)^s}{2}\frac{4l(l+1)(2l+1)}{\left(l^2+l+s^2\right)}\left(\frac{\Gamma\left(1+l+s\right) \Gamma \left(1+l-s\right)}{\Gamma \left(1+2l\right)\Gamma \left(2l+2\right)}\right),
\end{equation}
and
\begin{align}
\,_{s}\Tilde{k}^{(1)}_{lm}=\frac{(-1)^s}{2}&\frac{\Gamma\left(1+l+s\right) \Gamma \left(1+l-s\right)}{\Gamma \left(1+2l\right)\Gamma \left(2l+2\right)}\nonumber
\\
&\times\Bigg[-\frac{4lm(l+1)(2l+1)}{\left(l^2+l+s^2\right)}\left\{\frac{N}{M\omega}\left(2\psi(1+2l)+2\psi(2+2l)-\psi(1+l+s)-\psi(1+l-s)\right) \right\}
\nonumber
\\
& +8m^2 \left\{\psi(1+l+s)-\psi(1+l-s)-\frac{\pi m \left(l^2+l+s^2\right)}{l(l+1)(2l+1)}\right\}
+
\frac{8l(l+1)(2l+1)^2}{\left(l^2+l+s^2\right)}\Bigg]\,.
\end{align}
\subsection{Tidal Love numbers of an extremal Kerr black hole for different tidal perturbations}
In this Appendix, we will list the values $\,_{s}\Tilde{k}^{(0)}_{lm}$ and $\,_{s}\Tilde{k}^{(1)}_{lm}$ of TLNs for scalar, electromagnetic, and gravitational tidal perturbations:
\begin{itemize}
    \item {\bf Scalar perturbation} $\bm{(s=0)}$
    \begin{subequations}\label{eq:Extremal_scalar_s_0}
\begin{equation}
    \,_{0}\Tilde{k}^{(0)}_{lm} = -2(2l+1)\left(\frac{\Gamma\left(1+l\right) \Gamma \left(1+l\right)}{\Gamma \left(1+2l\right)\Gamma \left(2l+2\right)}\right),
\end{equation}
and
\begin{align}
\,_{0}\Tilde{k}^{(1)}_{lm}=\frac{4\Gamma\left(1+l\right) \Gamma \left(1+l\right)}{\Gamma \left(1+2l\right)\Gamma \left(2l+2\right)}
&\Bigg[-m(2l+1)\left\{\frac{N}{M\omega}\left(\psi(1+2l)+\psi(2+2l)-\psi(1+l)\right) \right\}
\nonumber
\\
& -\frac{\pi m^3}{(2l+1)}
+
(2l+1)^2\Bigg]\,,
\end{align}
where
\begin{equation}
N =  \frac{2mM\omega}{(2l+1)}~.
\end{equation} 
\end{subequations}

    \item {\bf Electromagnetic perturbation} $\bm{(s=+1)}$
\begin{subequations}\label{eq:Extremal_EM_s_+1}
\begin{equation}
    \,_{1}\Tilde{k}^{(0)}_{lm} = \frac{1}{2}\frac{4l(l+1)(2l+1)}{\left(l^2+l+1\right)}\left(\frac{\Gamma\left(2+l\right) \Gamma \left(l\right)}{\Gamma \left(1+2l\right)\Gamma \left(2l+2\right)}\right),
\end{equation}
and
\begin{align}
\,_{1}\Tilde{k}^{(1)}_{lm}=-\frac{1}{2}&\frac{\Gamma\left(2+l\right) \Gamma \left(l\right)}{\Gamma \left(1+2l\right)\Gamma \left(2l+2\right)}\nonumber
\\
&\times\Bigg[-\frac{4lm(l+1)(2l+1)}{\left(l^2+l+1\right)}\left\{\frac{N}{M\omega}\left(2\psi(1+2l)+2\psi(2+2l)-\psi(2+l)-\psi(l)\right) \right\}
\nonumber
\\
& +8m^2 \left\{\psi(2+l)-\psi(l)-\frac{\pi m \left(l^2+l+1\right)}{l(l+1)(2l+1)}\right\}
+
\frac{8l(l+1)(2l+1)^2}{\left(l^2+l+1\right)}\Bigg]\,,
\end{align}
where
\begin{equation}
N =  \frac{2mM\omega\left(l^2+l+1\right)}{l(l+1)(2l+1)}~.
\end{equation} 
\end{subequations}

\item {\bf Electromagnetic perturbation} $\bm{(s=-1)}$
\begin{subequations}\label{eq:Extremal_EM_s_-1}
\begin{equation}
    \,_{-1}\Tilde{k}^{(0)}_{lm} = \frac{1}{2}\frac{4l(l+1)(2l+1)}{\left(l^2+l+1\right)}\left(\frac{\Gamma\left(l\right) \Gamma \left(2+l\right)}{\Gamma \left(1+2l\right)\Gamma \left(2l+2\right)}\right),
\end{equation}
and
\begin{align}
\,_{-1}\Tilde{k}^{(1)}_{lm}=-\frac{1}{2}&\frac{\Gamma\left(l\right) \Gamma \left(2+l\right)}{\Gamma \left(1+2l\right)\Gamma \left(2l+2\right)}\nonumber
\\
&\times\Bigg[-\frac{4lm(l+1)(2l+1)}{\left(l^2+l+1\right)}\left\{\frac{N}{M\omega}\left(2\psi(1+2l)+2\psi(2+2l)-\psi(l)-\psi(2+l)\right) \right\}
\nonumber
\\
& +8m^2 \left\{\psi(l)-\psi(2+l)-\frac{\pi m \left(l^2+l+1\right)}{l(l+1)(2l+1)}\right\}
+
\frac{8l(l+1)(2l+1)^2}{\left(l^2+l+1\right)}\Bigg]\,,
\end{align}
where
\begin{equation}
N =  \frac{2mM\omega\left(l^2+l+1\right)}{l(l+1)(2l+1)}~.
\end{equation} 
\end{subequations}

\item {\bf Gravitational perturbation} $\bm{(s=+2)}$
\begin{subequations}\label{eq:Extremal_G_s_+2}
\begin{equation}
    \,_{2}\Tilde{k}^{(0)}_{lm} = -\frac{1}{2}\frac{4l(l+1)(2l+1)}{\left(l^2+l+4\right)}\left(\frac{\Gamma\left(3+l\right) \Gamma \left(l-1\right)}{\Gamma \left(1+2l\right)\Gamma \left(2l+2\right)}\right),
\end{equation}
and
\begin{align}
\,_{2}\Tilde{k}^{(1)}_{lm}=\frac{1}{2}&\frac{\Gamma\left(3+l\right) \Gamma \left(l-1\right)}{\Gamma \left(1+2l\right)\Gamma \left(2l+2\right)}\nonumber
\\
&\times\Bigg[-\frac{4lm(l+1)(2l+1)}{\left(l^2+l+4\right)}\left\{\frac{N}{M\omega}\left(2\psi(1+2l)+2\psi(2+2l)-\psi(3+l)-\psi(l-1)\right) \right\}
\nonumber
\\
& +8m^2 \left\{\psi(3+l)-\psi(l-1)-\frac{\pi m \left(l^2+l+4\right)}{l(l+1)(2l+1)}\right\}
+
\frac{8l(l+1)(2l+1)^2}{\left(l^2+l+4\right)}\Bigg]\,,
\end{align}
where
\begin{equation}
N =  \frac{2mM\omega\left(l^2+l+4\right)}{l(l+1)(2l+1)}~.
\end{equation} 
\end{subequations}

\item {\bf Gravitational perturbation} $\bm{(s=-2)}$
\begin{subequations}\label{eq:Extremal_G_s_-2}
\begin{equation}
    \,_{-2}\Tilde{k}^{(0)}_{lm} = -\frac{1}{2}\frac{4l(l+1)(2l+1)}{\left(l^2+l+4\right)}\left(\frac{\Gamma\left(l-1\right) \Gamma \left(3+l\right)}{\Gamma \left(1+2l\right)\Gamma \left(2l+2\right)}\right),
\end{equation}
and
\begin{align}
\,_{-2}\Tilde{k}^{(1)}_{lm}=\frac{1}{2}&\frac{\Gamma\left(l-1\right) \Gamma \left(3+l\right)}{\Gamma \left(1+2l\right)\Gamma \left(2l+2\right)}\nonumber
\\
&\times\Bigg[-\frac{4lm(l+1)(2l+1)}{\left(l^2+l+4\right)}\left\{\frac{N}{M\omega}\left(2\psi(1+2l)+2\psi(2+2l)-\psi(l-1)-\psi(3+l)\right) \right\}
\nonumber
\\
& +8m^2 \left\{\psi(l-1)-\psi(3+l)-\frac{\pi m \left(l^2+l+4\right)}{l(l+1)(2l+1)}\right\}
+
\frac{8l(l+1)(2l+1)^2}{\left(l^2+l+4\right)}\Bigg]\,,
\end{align}
where
\begin{equation}
N =  \frac{2mM\omega\left(l^2+l+4\right)}{l(l+1)(2l+1)}~.
\end{equation} 
\end{subequations}

\end{itemize}
\subsection{Tidal response of an extremal Kerr black hole in an axisymmetric tidal field}\label{app_B.3}

In this Appendix, we shall study the tidal response function of an extremal Kerr BH in the axisymmetric tidal background ($m=0$), separately. The tidal response function of an extremal Kerr BH, i.e., \ref{non_static_rf_ext_BH}, in the axisymmetric tidal field is
\begin{equation}\label{resp_func_ext_BH_axi_sym_11}
\,_{s}F_{lm}=-(-4iM\omega)^{2l+1}\frac{\Gamma \left(-2l\right) \Gamma \left(1+l-s-2 i M \omega\right)}{\Gamma \left(-l-s-2 i M \omega\right)\Gamma \left(2l+2\right)}\,.
\end{equation}
Similar to the previous Appendixes, we can expand the ratio of Gamma function in the linear orders of $M\omega$ and neglect the second- and higher-order terms in $M\omega$. This yields
\begin{equation}\label{simplify_ext_BH_axi_sym_1}
\,_{s}F_{lm}=-(-4iM\omega)^{2l+1}\frac{\Gamma \left(-2l\right) \Gamma \left(1+l-s\right)}{\Gamma \left(-l-s\right)\Gamma \left(2l+2\right)}\left[1-2iM\omega\,\psi(1+l-s)+2iM\omega\,\psi(-l-s)\right]\,.
\end{equation}
Since we are working the linear orders of $M\omega$ and there is a term of order $(M\omega)^{2l+1}$ outside the above expression, we can ignore the linear order terms of $M\omega$ inside the square brackets. In addition, for $l\in \mathbb{Z}_{\ge0}$, $s\in\mathbb{Z}$, and $l\ge |s|$, we can rewrite the tidal response function as
\begin{equation}\label{simplify_ext_BH_axi_sym_11}
\,_{s}F_{lm}=-\frac{(-1)^{s-l}}{2}(-4iM\omega)^{2l+1}\frac{\Gamma \left(1+l-s\right) \Gamma \left(1+l+s\right)}{\Gamma\left(2l+1\right)\Gamma\left(2l+2\right)}\,.
\end{equation}
Now, it is clear that the above expression is purely imaginary in nature due to the $i^{2l+1}$ as an overall multiplication factor. Thus, the tidal Love numbers of an extremal Kerr black hole in the axisymmetric tidal field are zero.

\bibliographystyle{apsrev4-1}
\bibliography{reference}
\end{document}